\documentclass[aps,prb,twocolumn,superscriptaddress]{revtex4-2}
\usepackage{amsmath,amssymb,amsfonts,upgreek}
\usepackage{color}
\usepackage{graphicx}
\usepackage{setspace}
\usepackage{bm}
\usepackage[export]{adjustbox}
\usepackage{xr-hyper}
\usepackage{hyperref}
\usepackage{dsfont}
\usepackage{braket}
\usepackage{siunitx}
\usepackage{soul}
\usepackage{changes}

\newcommand{\ttg}{\ensuremath{t_{2g}}}

\newcommand{\ttgmodel}{\ensuremath{t_{2g}}-\ensuremath{t_{2g}}}
\newcommand{\pddmodel}{\ensuremath{pd}-\ensuremath{d}}

\newcommand{\smo}{SrMoO$_3$}
\newcommand{\rfour}{$R_4^+$}
\newcommand{\cubic}{$Pm\overline{3}m$}
\newcommand{\ucrpa}{$U_\text{cRPA}$}

\begin{document}

\title{Correlation-Induced Octahedral Rotations in SrMoO$_3$}

\author{Alexander Hampel}
\email{ahampel@flatironinstitute.org}
\affiliation{Center for Computational Quantum Physics, Flatiron Institute, 162 Fifth avenue, New York, NY 10010, USA}
\author{Jeremy Lee-Hand}
\affiliation{Department of Physics and Astronomy,
             Stony Brook University,
             Stony Brook, New York, 11794-3800, USA}

\author{Antoine Georges}
\affiliation{Coll{\`e}ge de France, 11 place Marcelin Berthelot, 75005 Paris, France}
\affiliation{Center for Computational Quantum Physics, Flatiron Institute, 162 Fifth avenue, New York, NY 10010, USA}
\affiliation{CPHT, CNRS, {\'E}cole Polytechnique, IP Paris, F-91128 Palaiseau, France}
\affiliation{DQMP, Universit{\'e} de Gen{\`e}ve, 24 quai Ernest Ansermet, CH-1211 Gen{\`e}ve, Suisse}
\author{Cyrus E. Dreyer}
\affiliation{Department of Physics and Astronomy,
             Stony Brook University,
             Stony Brook, New York, 11794-3800, USA}
\affiliation{Center for Computational Quantum Physics, Flatiron Institute, 162 Fifth avenue, New York, NY 10010, USA}

\date{\today}

\begin{abstract}
Distortions of the oxygen octahedra influence the fundamental electronic structure of perovskite oxides, such as their bandwidth and exchange interactions.
Utilizing a fully \textit{ab-initio} methodology based on density functional theory plus dynamical mean field theory (DFT+DMFT), we study the crystal and magnetic structure of \smo{}. 
Comparing our results with DFT+$U$ performed on the same footing, we find that DFT+$U$ overestimates the propensity for magnetic ordering, as well as the octahedral rotations, leading to a different ground state structure. This demonstrates that structural distortions can be highly sensitive to electronic correlation effects, and to the considered magnetic state, 
even in a moderately correlated metal such as \smo{}.
Moreover, by comparing different downfolding schemes, we demonstrate the robustness of the DFT+DMFT method for obtaining structural properties, highlighting its versatility for applications to a broad range of materials.

\end{abstract}

\maketitle

\section{Introduction}

\begin{figure}[t]
    \centering
    \includegraphics[width=0.9\linewidth]{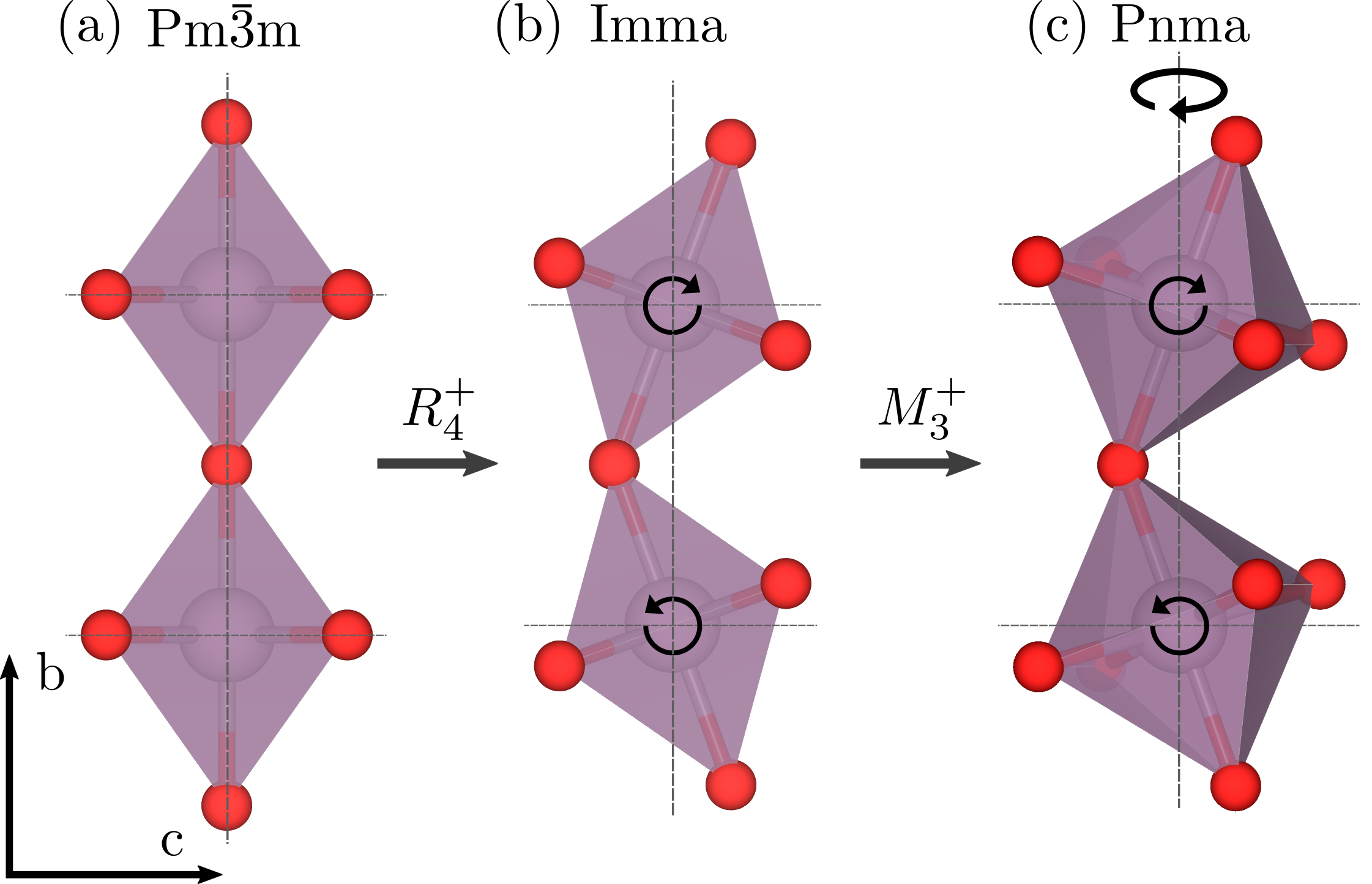}
    \caption{Schematic depiction of a common symmetry-lowering structural distortions found in perovskites. (a) Undistorted perovskite in the cubic space group \cubic{}. (b) Out-of-phase tilts of BO$_6$ octahedra along the $c$ axis lower the symmetry to $Imma$ (c) Additional BO$_6$ rotation around the $c$ axis results in the $Pnma$ space group. 
    }
    \label{fig:smo_struc}
\end{figure}

$AB$O$_3$ perovskite oxides exhibit a variety of exotic and technologically interesting phenomena including high-temperature superconductivity~\cite{RevModPhys.66.763}, non-Fermi liquid behavior~\cite{RevModPhys.73.797}, multiferroicity~\cite{Khomskii2009}, strong electron-lattice coupling~\cite{Haule2018}, and metal-insulator transitions (MIT's)~\cite{Imada1998}. The key to predicting such phases is a quantitative understanding of the relative importance of, e.g., strong electron correlations, spin-orbit coupling, magnetic properties, and connected structural distortions. In quantum materials, these may occur at similar energy scales, requiring \textit{ab-initio} theoretical approaches to simultaneously describe multiple phenomena with a high level of accuracy.

In perovskite oxides, the type and degree of rotations of the $B$O$_6$ octahedra (Fig.~\ref{fig:smo_struc}) is fundamentally tied to the electronic structure~\cite{Guzman-Verri2019, Cammarata2014}, as it determines the relative importance of the kinetic energy versus electron-electron interactions~\cite{Imada1998}.
Most often, density-functional theory (DFT), possibly including an empirical or \textit{ab-initio} chosen Hubbard $U$ interaction (DFT+$U$), is relied on to provide the general structural properties of perovskite oxides, i.e., those not \textit{a priori} associated with strong electron-electron interactions~\cite{Haule2018, leonov2008, peil:2019, Park2014short}.
Though quite successful in many materials~\cite{Balachandran2018}, an accurate electronic structure is necessary in general for correct structural predictions~\cite{Cammarata2014, Guzman-Verri2019,PhysRevB.85.054417}, which may require going beyond DFT/DFT+$U$~\cite{Haule2018, peil:2019, Varignon2019,Ramberger2017, leonov2008}.  

By considering the case of \smo{} (SMO), we show in this work that this may be the case even for materials with moderate correlations. Specifically, we analyze the sensitivity of the octahedral rotations to electronic correlation effects. To this end we first perform structural calculations for SMO on a static mean-field level using DFT+$U$, which we then compare to calculations including the dynamic correlations via combining DFT with dynamical mean-field theory (DMFT)~\cite{Kotliar:2006,held2007}. By calculating the effective screened Coulomb interaction by means of the constrained random phase approximation (cRPA) we perform \textit{ab-initio} calculations parameter free and compare DFT+$U$ and DFT+DMFT on the same footing. Thereby, we demonstrate that octahedral rotations  can be very sensitive to correlations, even if the considered material shows no electronic phase transition induced by the correlation effects. 

SMO is found experimentally to be cubic (\cubic) at room temperature, undergoing transitions to $I4/mcm$ at \SI{266}{K}, and to $Imma$ at \SI{124}{K}.~\cite{Ikeda2000, Macquart2010}. The structural transitions are characterized by a gradual increase of rotation angle, see Fig.~\ref{fig:smo_struc}b, upon cooling~\cite{Macquart2010}.
SMO is one of the best conducting materials among transition-metal oxides, with reported resistivities as low as 
$5\,\mu\Omega\mathrm{cm}$ at room temperature~\cite{Nagai2005}, an interesting property for possible applications to electronics. 
SMO is a Pauli paramagnet, with no reported magnetic ordering down to 2~K~\cite{Ikeda2000}.
These observations, as well as specific heat measurements yielding a quasiparticle mass renormalization 
$m^*/m_b\simeq 2$~\cite{Nagai2005, Wadati2014}, hint at a moderate degree of electronic correlations. 
However, previous DFT+$U$ studies show discrepancies with experimental structural and magnetic properties~\cite{Somia2019, Zhu2012, Tariq2018, LeeHand2020}, hinting at the role of correlations in the structural properties.

Here we will compare the lattice energetics of SMO between DFT, DFT+$U$, and DFT+DMFT using a symmetry-based mode decomposition~\cite{PerezMato:2010ix, Campbell:2006}. This allows us to test the relative stability of the high-temperature cubic, the low-temperature orthorhombic $Imma$, and the orthorhombic $Pnma$ structure (recently found to be lowest energy by DFT+$U$ \cite{LeeHand2020}) by systematically varying different symmetry-allowed lattice distortions. The reported $Imma$ and $I4/mcm$ structures~\cite{Macquart2010} are almost identical besides the gradual increase of the rotation angle. Both structures differ only in a marginal change in lattice constant mismatch $a \neq b$ of  $\approx 0.3 \%$~\cite{Macquart2010} in $Imma$ compared to $I4/mcm$ where $a=b$. Furthermore, in $Imma$ a very small octahedral distortion ($R_5^+$ mode) is found. Hence, we discuss here only the $Imma$ structure.

The rest of the article is organized as follows. In Sec.~\ref{sec:theo} we present the theoretical framework. Then, in Sec.~\ref{sec:results} we present our results, split into five parts. We analyze first in Sec.~\ref{sec:struc_dft} the structural predictions obtained by DFT and DFT+$U$, then we discuss in Sec.~\ref{sec:downfolding} the downfolding and screening, leading in Sec.~\ref{sec:spectral_prop} to the analysis of the spectral properties. In the last two parts of the results Sec.~\ref{sec:r4_dmft} and Sec.~\ref{sec:Imma_to_Pnma} we discuss the DFT+DMFT structural properties. We end with a brief summary in Sec.~\ref{sec:summary}.

\section{Theoretical Framework}
\label{sec:theo}

\subsection{DFT+$U$}

DFT calculations are performed using the projector augmented wave (PAW) method~\cite{Blochl:1994dx}, implemented in the Vienna Ab initio Simulation Package (VASP)~\cite{Kresse:1993bz,Kresse:1996kl,Kresse:1999dk}, and the exchange-correlation functional of Perdew, Burke, and Ernzerhof~\cite{Perdew:1996iq}. 
For the \smo{} DFT calculations we treated the following valence states explicitly: Sr ($4s,5s,4p,4d$), Mo ($4s, 5s, 4p, 4d, 4f$), and O ($2s, 2p, 3d$). For calculations in the cubic \cubic{} cell we used a $k$-point mesh with $15 \times 15 \times 15$ grid points along the three reciprocal lattice directions, whereas for the larger orthorhombic unit cells we used $9 \times 9 \times 7$ $k$-point grid throughout all calculations including the charge self-consistent (CSC) DFT+DMFT calculations. A plane wave energy cut-off of \SI{550}{eV} was used in all calculations, except for the phonon calculations, where a higher cut-off of \SI{1000}{eV} was necessary for convergence. Forces and stress were computed with a precision down to 10$^{-4}$~eV/\r{A}.
To account for the local Coulomb interaction in the Mo $d$ shell on a static mean-field level we add an effective on-site interaction $U$ and Hund's rule exchange interaction $J$ according to Ref.~\cite{Liechtenstein:1995ip}. 

For phonon calculations the frozen-phonon method as implemented in \textsc{phonopy}~\cite{phonopy} is utilized with a $2 \times 2 \times 2$ $q$-point grid. 

For the symmetry-based mode decomposition~\cite{PerezMato:2010ix} we use the software \textsc{ISODISTORT}~\cite{Campbell:2006}, where we normalize all distortion modes with respect to the pseudo-cubic parent structure (\textit{Ap} amplitudes). Moreover, we choose a unit cell setting with the Mo atom at the center of the cell. Within the experimentally observed $Imma$ structure only two distortion modes are allowed, the \rfour{} mode describing an octahedral rotation as shown in Fig.~1(b), and the mode $R_5^+$ which describes a bending of the O-Mo-O in-plane bonds. The latter one is found to be negligible in the experimental structure. Going from the $Imma$ to the $Pnma$ space-group allows for additional distortions. Most striking are the octahedral rotation mode $M_3^+$ [Fig.~\ref{fig:smo_struc}(c)] and the shearing mode $X_5^+$. More details of the modes found in DFT+$U$ can be found in Ref.~\onlinecite{LeeHand2020}. For our calculations in Fig.~2 and Fig.~4, we fix the lattice parameters to the ones provided by experiment, satisfied by the fact that the pseudocubic volume changes by less than 1\% from 300~K to 5~K~\cite{Macquart2010}.

\subsection{DFT+DMFT}
\label{sec:theo_dmft}

To perform DFT+DMFT calculations, we construct a correlated subspace by performing projections using PAW projectors~\cite{Amadon:2008} in VASP~\cite{Schuler_Aichhorn:2018} (i.e., the same projectors used in DFT+$U$), and utilize the interface to the \textsc{TRIQS/DFTTools} software package~\cite{aichhorn_dfttools_2016,parcollet_triqs_2015} and the soliDMFT~\cite{soliDMFT} software. We compare two different choices of correlated subspaces to underline consistency of the approach. First, we consider a minimal subspace model (labeled \ttgmodel{}), where we only construct local orbital projections related to the three Mo \ttg{} orbitals at the Fermi level from an energy window containing only these \ttg{} bands. Second, we construct the subspace using a large energy window model containing all O $2p$ and Mo $4d$ orbitals; Wannier functions are projected on all Mo $4d$ orbitals (labeled \pddmodel{}), which is comparable to the correlated subspace used in DFT+$U$, allowing a direct comparison.

In the \pddmodel{} we find an occupation of $\sim 4$ electrons in the $d$ shell due to the mixing with O $2p$ states. During the DMFT calculation the impurity occupation changes by less than 0.05 electrons, depending on the rotation amplitude. In the \ttgmodel{} model we find an occupation of exactly 2 electrons, i.e., the nominal occupancy of the Mo $4d$ state. This is because, in this minimal subspace, a unitary transformation connects local orbitals and Kohn-Sham states. 

The resulting effective impurity problem within the DMFT cycle is solved with a continuous-time QMC hybridization-expansion solver~\cite{Gull:2011} (cthyb) implemented in \textsc{TRIQS/cthyb}~\cite{Seth2016274}, taking into account all off-diagonal elements of the local Green's function in the crystal-field basis. We add a local Coulomb interaction in the form of the Hubbard-Kanamori Hamiltonian including all spin-flip and pair-hopping terms~\cite{vaugier2012} for the \ttgmodel{} and a density-density only interaction for the \pddmodel{} model with parameters obtained from cRPA. All calculations are performed fully charge self-consistent.


To optimize the sign, we rotate into the orbital basis which diagonalizes the impurity occupations. However, a treatment beyond density-density seems to be not feasible for the five orbital \pddmodel{} model; the sign problem is severe for calculations with octahedral rotations, as the hybridization functions develops off-diagonal elements. To correct the electron-electron interaction within the correlated subspace already accounted for within VASP, we use the fully-localized limit DC correction scheme~\cite{Solovyev:1994,anisimov1997} using the DMFT impurity occupations. For the \ttgmodel{} we use the adapted form given in Ref.~\onlinecite{held2007}. Within our frontier-bands model the DC potential acts only as a trivial shift that can be absorbed in the chemical potential, thus, not influencing the important charge transfer energy between O $2p$ and Mo $4d$ states, which we fix to the DFT provided value~\cite{Hampel2020}. As seen in Fig.~\ref{fig:DMFT_Aw}, both models give a very similar spectral function around the Fermi level, showing that our chosen DC scheme for the \pddmodel{} model behaves very similar to the \ttgmodel{} model for the states close to the Fermi level. 

Total energies are calculated using the formula given in Ref.~\onlinecite{Amadon:2008}, where the impurity interaction energy is calculated as the expectation value of $\braket{\hat{H}_\text{int}}$. This is done by measuring the impurity density matrix $\hat{\rho}^\text{imp}$ directly in the cthyb solver within the Fock basis
\begin{align}
\braket{\hat{H}_\text{int}} = \text{Tr}_\text{imp} \left[ \hat{\rho}^\text{imp} \hat{H}_\text{int} \right] ,
\end{align}
where $\text{Tr}_\text{imp}$ sums over all impurity orbital and spin degrees. This procedure is free of the high-frequency noise of the impurity self-energy and allows for very accurate determination of the interaction energy~\cite{PhysRevLett.115.256402}; we estimate the error to be $\sim\SI{2}{meV}$. This reduces the error in the total energy significantly~\cite{PhysRevLett.115.256402}. We sample the energy over $\sim 20$ converged DMFT iterations to obtain errors in energy $<\SI{3}{meV}$. Convergence is reached when the standard error of the impurity occupation within the last 10 DMFT iterations is smaller than $2 \times 10^{-3}$. Here, we neglect all entropy terms to the energy for simplicity.

In all DMFT calculations the impurity problem is solved at a temperature of $\beta=40$~eV$^{-1} \approx 290$~K, except for the \pddmodel{} calculations in Fig.~4, where we used $\beta=20$~eV$^{-1}\approx 580$~K for increased numerical stability. 

\subsection{Screened Coulomb interaction}

To calculate the screened Coulomb interaction for our chosen correlated subspace we use the cRPA method as implemented in \textsc{VASP}~\cite{Merzuk2015}. I.e., we calculate the static part of the screened Coulomb interaction $U(\omega=0)$ by constructing maximally localized Wannier functions (MLWF) \cite{PhysRevB.80.155134} using \textsc{Wannier90}~\cite{Mostofi_et_al:2014}, following the ideas of Ref.~\cite{vaugier2012}.

To reflect our chosen DMFT subspace the \pddmodel{} model is constructed via MLWFs for all O $2p$, Mo $4d$, and Sr $3d$ states; we find that this produces an analogous set of local orbitals to the projected ones used in the DMFT~\cite{Park2015_wan}. In the \ttgmodel{} model, we construct three Wannier functions corresponding to the three \ttg{} orbitals at the Fermi level. For the frontier bands only model it has been shown that MLWF and projectors give the same results when used within DMFT~\cite{Hampel2020}. For the large energy window the constructed MLWFs show a very similar DOS compared to the projector DOS created in VASP. Thus, we are confident that the screening is calculated within very similar orbitals as used in the DMFT calculation. We note that there can be small differences between projectors and MLWF~\cite{Park2015_wan}, but we will assume that the relative error of the RPA approximation is more severe that the choice of basis. Furthermore, we tested several different choices to construct the MLWFs by varying the number of input bands, number of local orbitals, and tested the convergence of the disentanglement procedure to ensure that we obtained the most localized basis set for the Mo $4d$ orbitals possible. 

For the cRPA calculation we used a $k$-point grid of $7 \times 7 \times 5$ in the $Pm\bar{3}m$ cubic structure with $\sim300$ empty bands, and a grid of $5 \times 5 \times 3$ for the orthorhombic cells with $\sim 500$ empty bands. Both with an energy cut-off of 500~eV. The polarization function is evaluated within the MLWF basis. To extract symmetrized interaction parameters we spherically averaged the full four index interaction tensor assuming spherical symmetry allowing us to obtain parameters for the Hubbard-Kanamori Hamiltonian used for the \ttgmodel{} model, and Slater parameters for the \pddmodel{}~\cite{vaugier2012}. For the latter model we assumed $F^4/F^2=0.625$ for better comparability with DFT+$U$. When fitting the cRPA four index tensor directly to the three independent radial integrals $F^0,F^2,F^4$ we find a different ratio of $F^4/F^2=0.83$. However, values for $U$ and $J$ stay the same. Investigating the reason and implications of such change of the $F^4/F^2$ ratio is left open for future investigations. We used the static averaged interaction parameters from the cubic structure for almost all calculations, except for those calculations in which we explicitly used the full $U_{ijkl}(\omega=0)$ tensor. 

\section{Results}
\label{sec:results}

\subsection{Structural predictions from DFT and DFT+$U$}
\label{sec:struc_dft}

\begin{figure}[t]
    \centering
    \includegraphics[width=0.95\linewidth]{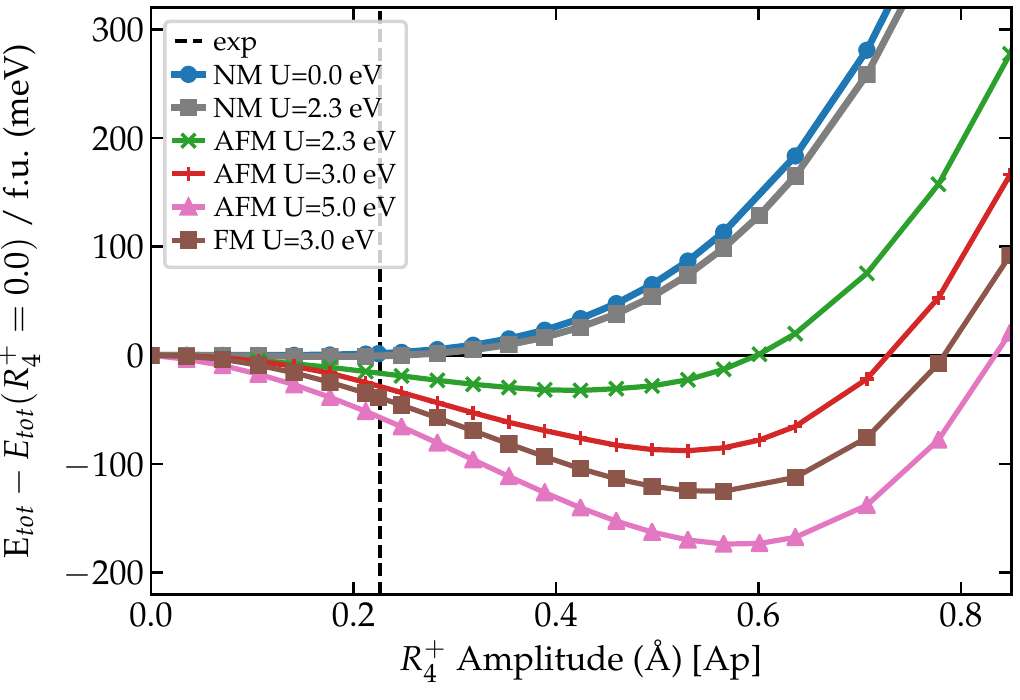}
    \caption{Energy versus \rfour{} octahedral rotation-mode amplitude for \smo{} calculated with DFT+$U$ (with fixed $J=\SI{0.7}{eV}$, except for $U$=0, where $J$=0). $U=\SI{2.3}{eV}$ corresponds to the value predicted by cRPA. The energy is given relative to the cubic structure, and the experimentally found \rfour{} amplitude is indicated by the dashed vertical line. Nonmagnetic (NM), antiferromagnetic (AFM) type-C, and ferromagnetic (FM) calculations are shown. 
    }
    \label{fig:DFT_energetics}
\end{figure}

\begin{figure}[t]
    \centering
    \includegraphics[width=0.75\linewidth]{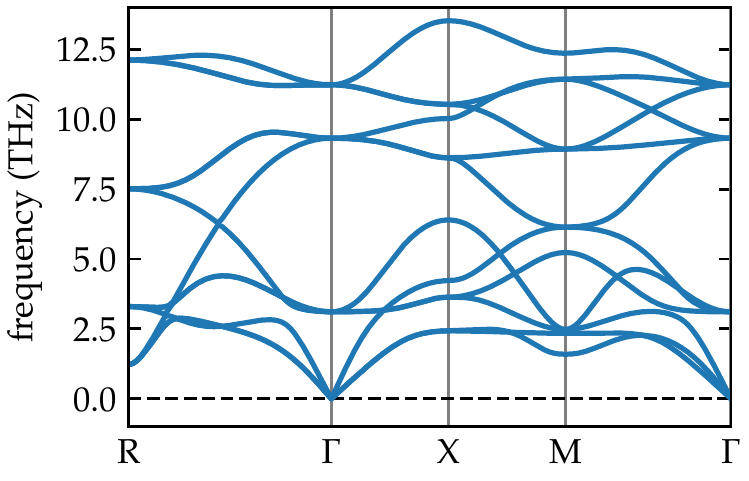}
    \caption{The \smo{} phonon dispersion in the cubic \cubic{} structure calculated with nonmagnetic DFT, showing no imaginary phonon modes.
    }
    \label{fig:smo_phonons}
\end{figure}

In Fig.~\ref{fig:DFT_energetics}, we plot the total energy, calculated with DFT, and DFT+$U$ (for different choices of Hubbard $U$) with respect to the amplitude of the \rfour{} octahedral rotation mode. This out-of-phase rotation of the MoO$_6$ octahedra around the $c$-axis takes $Pm\overline{3}m$, i.e., the experimental high-temperature structure of SMO, to $Imma$, the low-temperature phase~\cite{PerezMato:2010ix} [see Fig.~\ref{fig:smo_struc}(a) and (b)]. The energy is referenced to that of the cubic phase, i.e., with the \rfour{} amplitude set to zero, but using the experimental lattice parameters of $Imma$. The vertical dashed line is the experimental \rfour{} amplitude~\cite{Macquart2010}. DFT with $U=0$ eV predicts a nonmagnetic (NM) cubic structure, even when starting from a spin-polarized initial state for various orderings~\cite{LeeHand2020}; as we see in Fig.~\ref{fig:DFT_energetics}, increasing the \rfour{} amplitude only serves to increase the energy. There is a significant range of \rfour{} amplitudes where the energy changes very little, indicating that this mode is quite soft. This picture is confirmed by the phonons calculated with DFT in the NM state (see Fig.~A\ref{fig:smo_phonons}), which show no instabilities (modes of imaginary frequency), but a very soft \rfour{} mode with frequency of \SI{1}{Thz}. Clearly we must go beyond DFT to describe the low temperature structure of SMO.

\begin{table*}[t]
  \centering
    \caption{Screened Coulomb interaction values as calculated from cRPA in the static limit. The first three columns denote the model used for the Wannier construction, crystal structure, and the energy window. The rest of the columns show the cRPA results, including the bare Coulomb interaction $V$, screened orbitally-averaged Coulomb matrix elements in the Wannier basis $\mathcal{U}$, screening strength ratio of the two $\mathcal{U}/V$, Hund's exchange in the Wannier basis $\mathcal{J}$, and spherically averaged Slater parameter $U=F^0$ and $J=(F^2+F^4)/14$.}
\label{tab:cRPA-comp}
  \begin{ruledtabular}
\begin{tabular}{l c c c c c c c c}
\vspace{0.05cm}
model & structure &window~(eV) & $V$~(eV) & $\mathcal{U}$~(eV) & $\mathcal{U}/V$ & $\mathcal{J}$~(eV) & $U_\text{cRPA}=F^0$~(eV) &  $J$~(eV)  \\ \hline
\ttg - \ttg & \cubic{}       & [-2.5, 2.5] & 11.68 & 3.15 & 0.27 & 0.33 & $-$ &  $-$ \\ 
\ttg - \ttg & $Imma$ & [-2.5, 2.5] & 11.55 & 3.11 & 0.27 & 0.33 & $-$ & $-$ \\ 
\ttg - \ttg & $Pnma$ & [-2.5, 2.5] & 11.46 & 3.03 & 0.26 & 0.33 & $-$ & $-$ \\ 
$pd$-$d$ & \cubic{}     & [-9.0, 9.0] & 15.73 & 3.12 & 0.20 & 0.51 & 2.32 & 0.71  \\ 
\end{tabular}
\end{ruledtabular} 
\end{table*}

With the addition of $U$, magnetic order is stabilized \cite{LeeHand2020} (in contrast to experiments~\cite{Macquart2010, Ikeda2000}); we consider both ferromagnetic (FM) and $C$-type anti-ferromagnetic (AFM) order (which is lowest in energy~\cite{LeeHand2020,Tariq2018,Somia2019}). This leads to a stable orthorhombic structure, i.e., to a minimum at a finite \rfour{} mode amplitude. Thus, including an extra local Coulomb interaction on the Mo $4d$ states drives the system towards the correct structural phase. Note that suppressing the magnetic order by performing NM DFT+$U$ calculations (see gray squares in Fig.~\ref{fig:DFT_energetics}), gives similar results to the NM DFT calculations, i.e., the octahedral rotations are suppressed.

The experimental reported \rfour{} amplitude at \SI{5}{K} is \SI{0.23}{\angstrom}, which corresponds to a change in bond angle of ~4.4$^\circ$~\cite{Macquart2010}. For $U=\SI{3}{eV}$ and AFM order, we find $R_4^+=\SI{0.52}{\angstrom}$, thus octahedral rotations are more than twice as large compared to experiment. By systematically varying $U$ (keeping AFM order) in a reasonable regime, we find that increasing $U$ gives larger equilibrium $R_4^+$ amplitudes. For $U=\SI{5}{eV}$ we find $R_4^+=\SI{0.58}{\angstrom}$, whereas for a smaller $U=\SI{2.3}{eV}$ (which is the $U$ value predicted by cRPA, more details below) we find $R_4^+=\SI{0.43}{\angstrom}$. FM order results in slightly larger rotation amplitude compared to AFM order (Fig.~\ref{fig:DFT_energetics}). Also, when performing full structure optimizations, additional energy-lowering octahedral rotation modes are activated, leading to the $Pnma$ structure instead of $Imma$~\cite{LeeHand2020} [See Fig.~\ref{fig:smo_struc}(b) and (c)]. 
Overall, we see from Fig.~\ref{fig:DFT_energetics} that the stability of the orthorhombic structure and the \rfour{} amplitude are very sensitive to the Coulomb interaction in the Mo $4d$ orbitals in any magnetically ordered state, and are clearly overestimated by DFT+$U$. For NM calculations, even with a finite $U$, the \rfour{} rotations are completely suppressed. Thus, we also must go beyond DFT+$U$ to capture the correct ground-state structure for SMO.

\subsection{Downfolding \& cRPA}
\label{sec:downfolding}

\begin{figure}[t]
    \centering
    \includegraphics[width=0.9\linewidth]{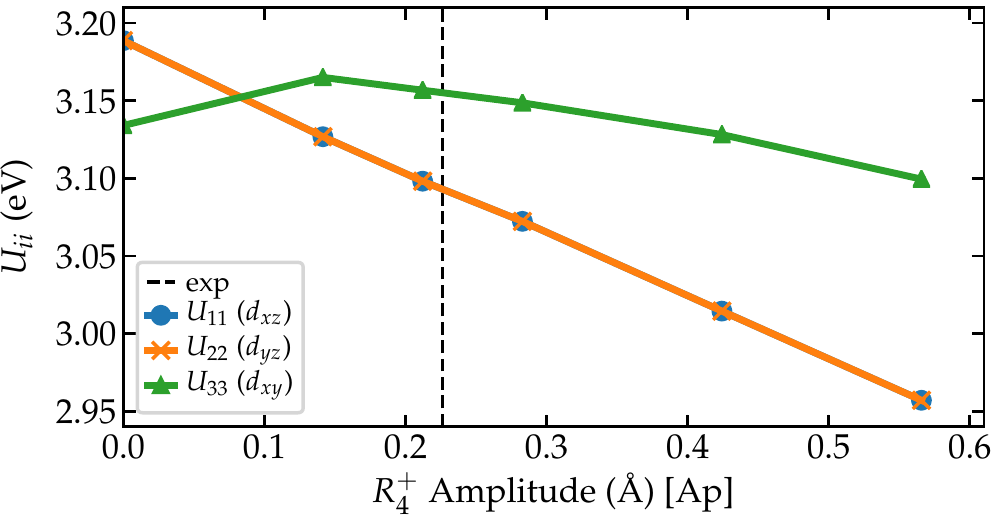}
    \caption{On-site Coulomb interaction tensor values $U_{iijj}(\omega=0)$ for the \ttgmodel{} model as function of the \rfour{} amplitude in the $Imma$ structure. The experimental \rfour{} amplitude is shown by a vertical dashed line.}
    \label{fig:crpa_r4}
\end{figure}

As mentioned in Sec.~\ref{sec:theo_dmft} we compare two different choices of correlated subspaces. The minimal \ttgmodel{} subspace model, where we only construct local orbital projections related to the three Mo \ttg{} orbitals at the Fermi level. Second, the large window \pddmodel{} containing all O $2p$ and Mo $4d$ orbitals. The projected density of states are shown in Fig.~\ref{fig:DMFT_Aw}(a). 
Comparing the two models allows us to understand the importance of the $e_g$ orbitals in the active subspace (although they are nominally unoccupied), and the hybridization to the ligand states.

To cross-check the potential influence of the rotations on the Coulomb interaction we performed a series of cRPA calculations for different \rfour{} amplitudes. The results are shown in Fig.~\ref{fig:crpa_r4} for the \ttgmodel. We find that the average $\mathcal{U}$ is only reduced by $\sim 5$\% from $R_4^+=\SI{0.0}{\angstrom}$ to $R_4^+=\SI{0.8}{\angstrom}$. Furthermore, the degeneracy between the onsite values is lifted slightly and differ maximally by 5\% for $R_4^+=\SI{0.8}{\angstrom}$.
Therefore, we use the orbitally-averaged Coulomb interaction values assuming $F^4/F^2=0.63$~\cite{vaugier2012}. The resulting effective parameters are shown in Table~\ref{tab:cRPA-comp}. 
This approach allows us to compare DFT+$U$ and DFT+DMFT directly, by using exactly the same form of interaction and the same downfolding procedure in both calculations. A detailed discussion about orbital dependent results, including the full Coulomb interaction matrices are presented in Appendix \ref{app:dn_folding}.

For the \ttgmodel{} our Hubbard-Kanamori parameters are in agreement with Refs.~\cite{Nilsson2017,Petocchi2020, vaugier2012} which find a value of $\mathcal{U}\sim \SI{3.1}{eV}$ using the same approach. 
Our parameters for the \pddmodel{} model are also given in Table~\ref{tab:cRPA-comp}, and are consistent with Ref~\cite{Petocchi2020}. We see that going from the small (\ttgmodel{}) to the large (\pddmodel{}) energy window does not increase in the Coulomb interaction, even though the orbitals are significantly more localized (reflected in the larger bare Coulomb interaction $V$ in Table~\ref{tab:cRPA-comp}, and the increased $\mathcal{J}$). Therefore including the $e_g$ orbitals in the active space results in more effective screening of the active subspace due to the the large $e_g$-O $2p$ hybridization~\cite{Petocchi2020}.

In our calculations of the \ttgmodel{} model we use the Hubbard-Kanamori form of the interaction Hamiltonian, including all spin-flip and pair-hopping terms~\cite{vaugier2012} parametrized by $\mathcal{U}$ and $\mathcal{J}$, and for the \pddmodel{} model we only include density-density type interactions parametrized by \ucrpa{} (i.e., $F^0$) and $J$; \ucrpa{} also serves as the cRPA prediction for the Coulomb interaction appropriate for DFT+$U$, as discussed above.

\subsection{DFT+DMFT: Spectral properties}
\label{sec:spectral_prop}

\begin{figure}[t]
    \centering
    \includegraphics[width=0.95\linewidth]{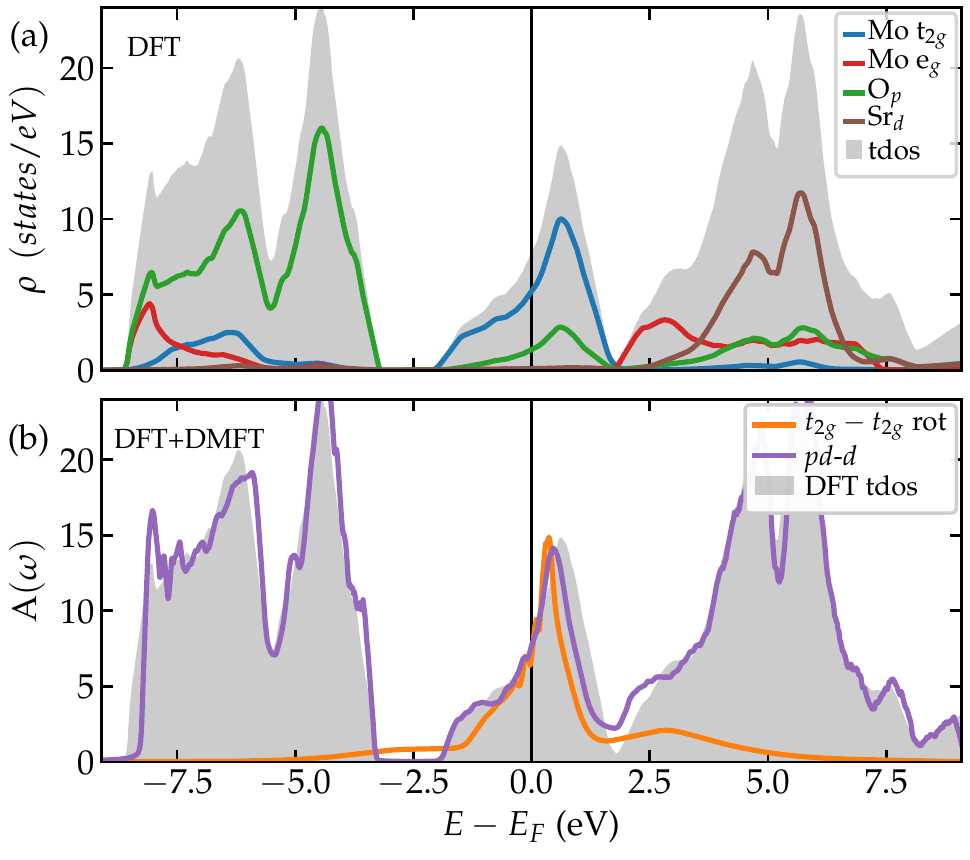}
    \caption{(a) Nonmagnetic (NM) DFT DOS and (b) DFT+DMFT spectral functions for the experimental low temperature $Imma$ structure. The \ttgmodel{} model (orange) and \pddmodel{} model (purple) using \ucrpa{} are both displayed in comparison to the total DFT DOS (gray).
    }
    \label{fig:DMFT_Aw}
\end{figure}

We first perform DFT+DMFT calculations for both correlated subspaces in the low-temperature experimental $Imma$ structure. 
As stated in Sec.~\ref{sec:theo_dmft} all calculations are performed above 0~K in DMFT. However, this temperature corresponds only to the electronic temperature of the system, and we extract the total energy to approximate the zero temperature limit. Furthermore, we checked that for \smo{} good quasiparticle behavior is found already at very high temperatures (see Fig.~\ref{fig:Z_tau_temp} in Appendix~\ref{app:quasiparticles}). Hence, we expect little temperature dependence on the spectral and structural properties in DFT+DMFT.

Fig.~\ref{fig:DMFT_Aw}(b) compares the resulting DFT+DMFT spectral function $A(\omega)$, obtained by analytical continuation of the self energy to the real frequency axis~\cite{Kraberger2017, Jarrel:2010}, with the total DOS from DFT. For both models, \ttgmodel{} and \pddmodel, we find a very similar renormalization of the \ttg{} states around the Fermi level, with a quasiparticle weight of $Z \approx 0.6$ for the \ttgmodel{} model ($Z \approx 0.7$ for the \pddmodel{} model). These values are consistent with previous DMFT studies on the high-temperature cubic structure~\cite{Wadati2014, Petocchi2020}, and indicate a moderately correlated metallic state. We observe a small renormalization within the O $2p$ states at \SI{-8}{eV} and of the lower end of the \ttg{} states for the \pddmodel{} model which can be not resolved in the \ttgmodel{} model used in earlier studies~\cite{Wadati2014}. However, both models show a very similar $p$-$d$ splitting to DFT. 

\begin{figure}[t]
    \centering
    \includegraphics[width=0.9\linewidth]{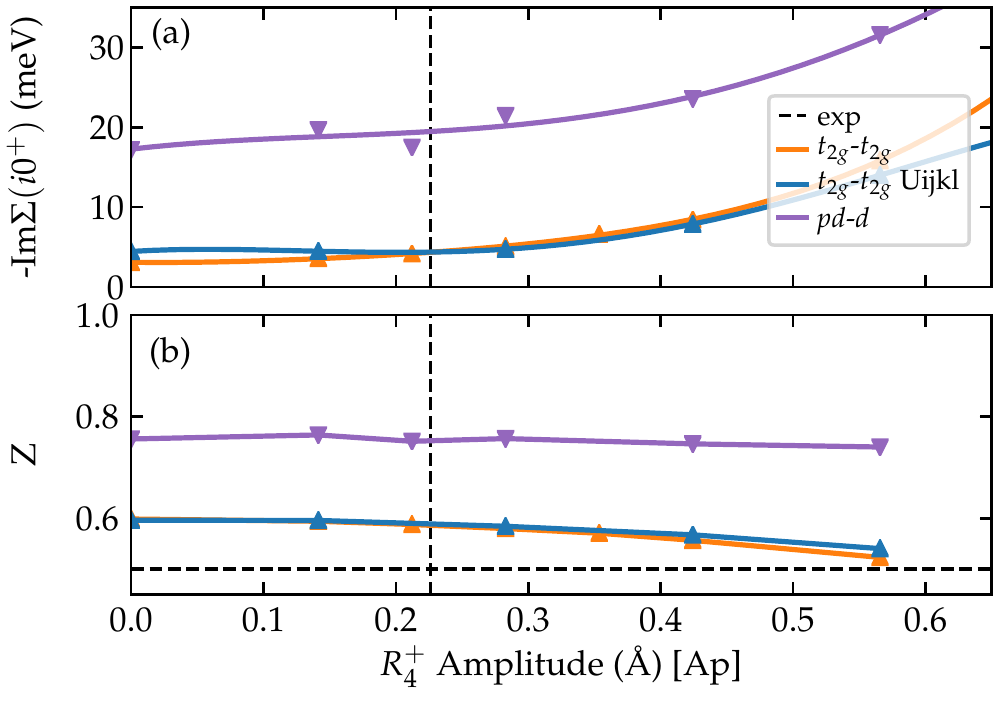}
    \caption{Comparison of (a) qasiparticle scattering rate -Im$\Sigma(i0^+)$ and (b) quasiparticle weight $Z$ as function of the \rfour{} amplitude of $Imma$ \smo{}. The blue triangles show the results when the full CSC DFT+DMFT calculation is performed using the full $U_{ijkl}$ tensor from cRPA. \ttgmodel{} calculations are performed at $T\approx$\SI{290}{K} and \pddmodel{} calculations at $T\approx$\SI{580}{K}.
    }
    \label{fig:Z_tau}
\end{figure}

We also extracted the scattering rate $- \text{Im} \Sigma(i 0^+)$~\cite{Xiaoyu2014,Ferber2012} for both models over a range of \rfour{} amplitudes. In Fig.~\ref{fig:Z_tau}, we display both $Z$ (Fig.~\ref{fig:Z_tau}(a)) and $-\mathrm{Im}\Sigma(i0^+)$ (Fig.~\ref{fig:Z_tau}(b)) as a function of the \rfour{} rotation amplitude. The scattering rate is as low as $\approx \SI{4}{meV}$ for the \ttgmodel{} model and $\approx \SI{20}{meV}$ for the \pddmodel{} model calculations at $R_4^+=0$ indicating long quasiparticle lifetimes $[Z\text{Im}\Sigma(i 0^+)]^{-1}$. See Appendix \ref{app:quasiparticles} for a detailed analysis of the temperature dependence, which is found to be consistent with the $T^2$ Fermi liquid behavior observed experimentally for this compound over a rather extended temperature range (see Fig.~A\ref{fig:Z_tau_temp}). 

Within DMFT, the effect of electron-electron interactions enters the direct-current conductivity through the scattering rate $\mathrm{Im}\Sigma(i0^+)$, which we explain in detail in Appendix \ref{app:quasiparticles}. As displayed in Fig.~\ref{fig:Z_tau}(b), this rate does not vary significantly for \rfour{} amplitudes between \SI{0.0}{\angstrom} and \SI{0.3}{\angstrom}. This is in agreement with the experimental fact that no significant drop in resistivity is observed at the structural transition~\cite{Nagai2005}. 

To check the influence of the full Coulomb interaction tensor on structural predictions, in contrast to the averaged values from the cubic structure used so far, we also performed DFT+DMFT calculations in the \ttgmodel{} model using the full $U_{ijkl}(\omega=0)$ tensor for each of the calculated structures. For the DC correction in those calculations we used the corresponding averaged interaction values. 
Using the full Coulomb interaction tensor from cRPA for each structure we find that the average quasiparticle properties $Z$ and $\mathrm{Im}\Sigma(i0^+)$ do not change (see Fig.~\ref{fig:Z_tau}), even though we observe small changes in the orbital dependent properties. 
%

\subsection{DFT+DMFT: octahedral rotations in $Imma$}
\label{sec:r4_dmft}

Now, we turn towards the dynamic stability of the $Imma$ phase within DFT+DMFT and compare to the DFT+$U$ results. DFT+DMFT total energy calculations~\cite{Amadon2006} are performed in the paramagnetic state varying the \rfour{} amplitude, keeping all other structural parameters fixed to experimental values. To obtain high-accuracy results, we sample the energy over several converged DMFT iterations and measure the interaction energy directly in the impurity solver via the impurity density matrix~\cite{PhysRevLett.115.256402}; we estimate the error in the energy to be $<\SI{3}{meV}$.  

\begin{figure}[t]
    \centering
    \includegraphics[width=0.95\linewidth]{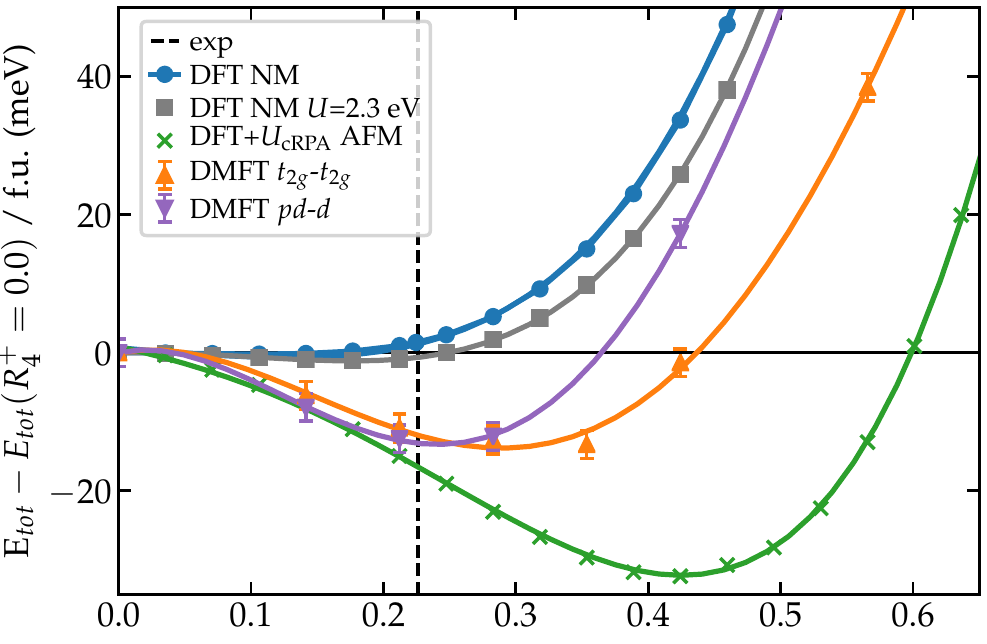}
    \caption{
    Energy versus \rfour{} octahedral rotation mode amplitude for \smo{} as calculated within DFT and DFT+DMFT, relative to the cubic structure. The experimental \rfour{} amplitude is indicated by the dashed vertical line. The NM DFT result (blue) and the DFT+\ucrpa{} result (green) are identical to Fig.~\ref{fig:DFT_energetics}. The DFT+DMFT results are shown for the \ttgmodel{} model (orange) and for the \pddmodel{} model (purple) using \ucrpa. Lines represent a 4th-order polynomial fit, and the error bars for DMFT are estimated to be \SI{2}{meV}. 
     }
    \label{fig:DMFT_energetics}
\end{figure}

Fig.~\ref{fig:DMFT_energetics} displays our results, where the NM DFT and AFM DFT+\ucrpa{} from Fig.~\ref{fig:DFT_energetics} are shown for comparison. First, we perform calculation for the minimal \ttgmodel{} model (orange line). The resulting total energy as function of the \rfour{} amplitude shows a clear minimum at around \SI{0.29}{\angstrom}, compared to DFT+\ucrpa{} at \SI{0.43}{\angstrom}. For the \pddmodel{} (purple curve in Fig.~\ref{fig:DFT_energetics}) we obtain an even better agreement to experiment with $R_4^+=\SI{0.24}{\angstrom}$. These values where obtained by performing a polynomial fit of fourth order to the data points, with very small error as seen in Fig~\ref{fig:DMFT_energetics}. Further, we emphasize that the \pddmodel{} model and the DFT+$U$ formalism use the very same projectors within VASP for the construction of the correlated subspace. Hence, results can be compared on a quantitative level, to elucidate the role of dynamic correlations, as well as the difference between the magnetically ordered state considered in DFT+$U$ and the paramagnetic state in DFT+DMFT. Furthermore this shows, that the different models \ttgmodel{} and \pddmodel{} result in very similar energetics.

We stress that these calculations are performed with completely analogous computational parameters as DFT+$U$, as we averaged the cRPA obtained Coulomb interaction tensors, thus allowing for a direct comparison between the different levels of theory. We find that DFT+$U$ gives considerably larger \rfour{} amplitudes, overestimating those in experiment, while DFT+DMFT results in rotations more comparable to experiment. Furthermore, the results depend on the chosen form of magnetic order (see Fig.~\ref{fig:DFT_energetics}). 

The DFT+DMFT calculations are performed within a truly paramagnetic state as observed in experiment~\cite{Macquart2010, Ikeda2000, Wadati2014}. In our calculations we did not observe any tendencies to form long-range magnetic order in DFT+DMFT. To this end we calculated the static spin-susceptibility down to \SI{40}{K} in DMFT, displaying a very small linear response (not shown), with no indications of long range order. As DMFT is known to overestimate ordering temperatures due to a lack of true spatial fluctuations we have significant confidence in the paramagnetic state predicted here~\cite{Dang2015,Kim2015,schaefer2020}.

When performing the total energy calculations with the full $U_{ijkl}(\omega=0)$ tensor (not shown) we observe a shift to slightly larger \rfour{} amplitudes closer to the DFT+\ucrpa{} (fixed cubic \ucrpa) predicted value. We observe that the non-symmetric interaction tensor reduces the interaction energy when the \rfour{} amplitude is increased compared to a symmetrized Kanamori interaction, which leads to a larger equilibrium \rfour{} amplitude. This is due to the fact that the averaged interaction energy decreases with increasing \rfour{} amplitude, and because the orbitals can reorganize occupations so that the interaction energy is minimized.
However, such results should be interpreted with caution, as an orbital dependent interaction would also need an appropriate orbital dependent DC scheme, for example the exact DC scheme proposed by \citeauthor{Haule:2015_exactDC}~\cite{Haule:2015_exactDC} or the orbital dependent corrections proposed by Ref.~\cite{Kristanovski:2018} and \cite{Nekrasov2012}, which is beyond the scope of this work. Note, that the question of DC is especially problematic as one has to compare small structural energy differences of the order of meV with large Coulomb interaction changes in the order 0.1~eV, that need to be correctly captured by the DC scheme. 
Parameterizing the interaction parameters fixes the dependence allowing for a better comparability between DFT+$U$ and DFT+DMFT, as it is also common practice when comparing structures in DFT+$U$.

\subsection{Comparison between $Imma$ and $Pnma$}
\label{sec:Imma_to_Pnma}

\begin{figure}[t]
    \centering
    \includegraphics[width=0.95\linewidth]{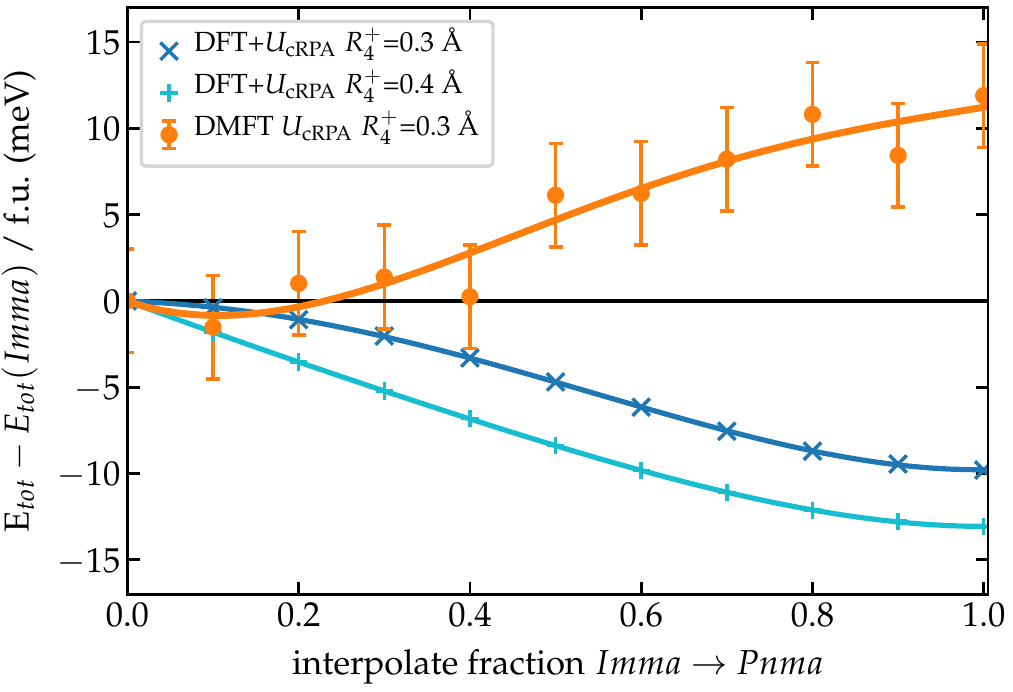}
    \caption{Relative energy of $Imma$ versus $Pnma$ structure for AFM DFT+\ucrpa{} and DFT+DMFT (\ttgmodel{} model) on a linearly interpolated path between the experimentally observed $Imma$ structure ($x=0$) and the AFM DFT+\ucrpa{} predicted $Pnma$ structure ($x=1$). The DFT+DMFT $Imma$ \rfour{} amplitude of \SI{0.3}{\angstrom} is kept fixed for DFT+DMFT (orange), and the additional $Pnma$ distortions are introduced according to the relaxed DFT+\ucrpa{} structure. The DFT+\ucrpa{} calculations are performed for fixed $R_4^+=\SI{0.3}{\angstrom}$ (blue) and $R_4^+=\SI{0.4}{\angstrom}$ (cyan), where the latter corresponds to the DFT+\ucrpa{} predicted value. Lines represent a polynomial fit, and the error bars for DMFT are estimated to be \SI{3}{meV}.
    }
    \label{fig:Imma_to_Pnma}
\end{figure}
As mentioned above, the AFM DFT+\ucrpa{} relaxation results in a $Pnma$ structure with additional octahedral rotations that are not observed in experiment \cite{Macquart2010}. To check whether DFT+DMFT correctly predicts the $Imma$ structure to be most stable, we perform calculations on linearly interpolated structures between the experimental $Imma$ and DFT+\ucrpa{} predicted $Pnma$ structure, while keeping lattice parameters constant. In practice this means that we fix $R_4^+$ amplitude and systematically introduce the additional $Pnma$ distortions $M_3^+$ and $X_5^+$ on top of the $Imma$ structure (see Fig.~\ref{fig:smo_struc}). For the DFT+\ucrpa{} calculations, we fix $R_4^+$ to the DFT+\ucrpa{} predicted value of \SI{0.4}{\angstrom} for $Imma/Pnma$, as well as the DFT+DMFT predicted value for $Imma$ ($R_4^+=\SI{0.3}{\angstrom}$); for DFT+DMFT (\ttgmodel{} model), we perform the calculation only for $R_4^+=\SI{0.3}{\angstrom}$.

The results are depicted in Fig.~\ref{fig:Imma_to_Pnma}. The DFT+\ucrpa{} calculations for both \rfour{} amplitudes (blue +'s and cyan x's) show a clear lowering of energy towards the $Pnma$ structure of \SI{10}{meV} to \SI{15}{meV} per formula unit compared to $Imma$. This is in agreement with recent results from Ref.~\cite{LeeHand2020}. In contrast, the DFT+DMFT result (orange circles) shows a clear increase of energy towards the $Pnma$ structure of about \SI{10}{meV}, predicting the $Imma$ structure to be lowest in energy in agreement with experiment. We note, that the energy accuracy in DFT+DMFT is not as good as in Fig.~\ref{fig:DMFT_energetics} as the impurity solver has to cope with small off-diagonal elements in the hybridization due to the additional distortions. Nevertheless, the data shows a very clear trend beyond the size of the estimated error of \SI{3}{meV}. 

We conclude that our treatment of the correlations on the level of DFT+DMFT predicts the crystal structure as well as the octahedral rotations for \smo{} consistent with experimental observations. This is a result of the calculations correctly capturing the paramagnetic state of the material and describing dynamic correlation effects.

\section{Summary}
\label{sec:summary}

We utilized a fully \textit{ab-initio} DFT+DMFT methodology in combination with symmetry-adapted distortion modes to accurately predict the level of octahedral rotations of \smo{} compared to experiment, while showing that DFT and DFT+$U$ compared on the same footing give drastically different results. We find that magnetic DFT+$U$ calculations, even when performed with exactly the same Coulomb interaction, give rise to significantly larger octahedral rotations. Thereby, we highlight the importance of correctly addressing the correlations and the paramagnetic state for structural predictions in \smo{}.
The tight coupling of the electronic and crystal structure in SMO is likely a result of the relatively flat potential energy surface from NM DFT with respect to \rfour{} octahedral rotations.
This work demonstrates that the structural properties of perovskite oxides can depend sensitively on the treatment of electron correlations, even when the structure is not obviously connected to a specific electronic phase transition, e.g., magnetic order, charge order, or MIT. 
Hence, showing that a quantitative understanding of the coupling between octahedral rotations and correlation effects is crucial for electronic structure engineering of perovskites, e.g. via heterostructuring or applying strain.

\begin{acknowledgments}
We thank Sophie D. Beck and Claude Ederer for insightful discussions about our results. CED and JLH acknowledge support from the National Science Foundation under Grant No.~DMR-1918455. 
AG acknowledges the support of the European Research Council (ERC-319286-QMAC).
The Flatiron Institute is a division of the Simons Foundation.
\end{acknowledgments}

\appendix
\section{Full U tensor from cRPA}
\label{app:dn_folding}

\begin{figure}
    \centering
    \includegraphics[width=\linewidth]{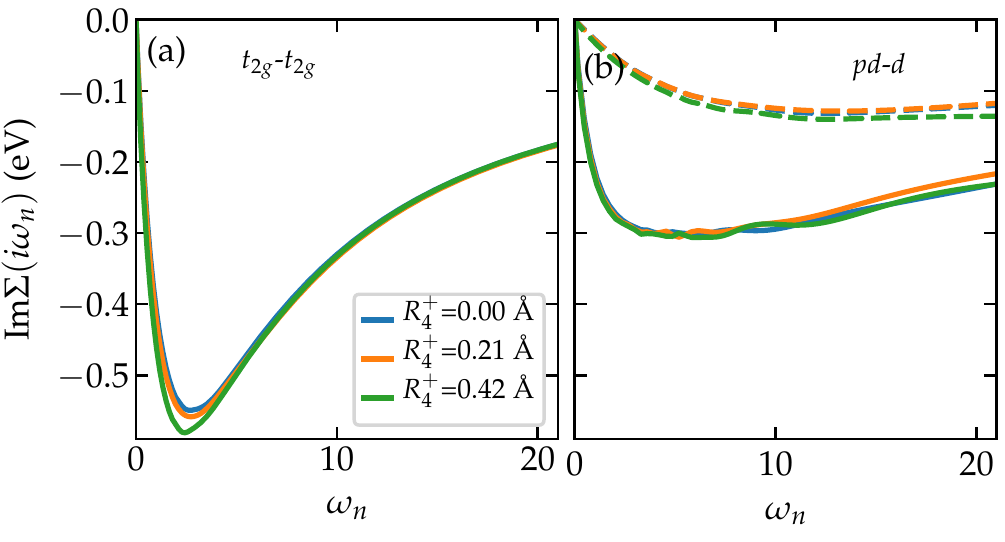}
    \caption{Imaginary part of the DMFT impurity Matsubara self-energies $\Sigma^{\text{imp}}(i \omega_n)$ calculated for different \rfour{} amplitudes, using the cRPA interaction values. (a) Imaginary part of $\Sigma^{\text{imp}}(i \omega_n)$ for the \ttgmodel{} model averaged over all three $t_{2g}$ orbitals. (b) Imaginary part of $\Sigma^{\text{imp}}(i \omega_n)$ for the \pddmodel{} model. Solid lines represent the averaged $t_{2g}$ orbitals, whereas dashed lines show the $e_g$ orbital averaged $\Sigma^{\text{imp}}(i \omega_n)$.
    }
    \label{fig:self_energies}
\end{figure}

For completeness we list here the averaged reduced screened interaction matrices for $\omega=0$ for parallel and anti-parallel spin for the \ttgmodel{} model (cubic, exp. $Imma$, and DFT+$U$ predicted $Pnma$ structure) and the \pddmodel{} model (cubic), which are defined as~\cite{vaugier2012}:
\begin{align}
    U_{ij}^{\sigma \bar{\sigma}} &= U_{ijij} \\
    U_{ij}^{\sigma \sigma} &= U_{ijij} - \underbrace{U_{ijji}}_{J_{ij}} \quad .
\end{align}
Here, $m=\{d_{xz}, d_{yz}, d_{xy}\}$ for the \ttgmodel{} model, and $m=\{d_{z^2},d_{xz}, d_{yz}, d_{x^2-y^2}, d_{xy}\}$ for the \pddmodel{} model. From cRPA we obtained for the \ttgmodel{}  model in the cubic structure (all values in eV):
\begin{align*}
    U_{ij}^{\sigma \bar{\sigma}} &=
    \begin{pmatrix}
  3.151 & 2.457 & 2.457\\
  2.457 & 3.151 & 2.457\\
  2.457 & 2.457 & 3.151\\
\end{pmatrix} \\
U_{ij}^{\sigma \sigma} &=
\begin{pmatrix}
  0. & 2.128 & 2.128\\
  2.128 & 0. & 2.128\\
  2.128 & 2.128 & 0.\\
\end{pmatrix}
\end{align*}
for the \ttgmodel model in experimental $Imma$ structure we obtained (all values in eV) :
\begin{align*}
    U_{ij}^{\sigma \bar{\sigma}} &=
    \begin{pmatrix}
  3.094 & 2.398 & 2.424\\
  2.398 & 3.094 & 2.424\\
  2.424 & 2.424 & 3.155\\
\end{pmatrix} \\
U_{ij}^{\sigma \sigma} &=
\begin{pmatrix}
 0. & 2.071 & 2.091\\
  2.071 & 0. & 2.091\\
  2.091 & 2.091 & 0.\\
\end{pmatrix}
\end{align*}
Here, we see that the changes from the cubic to the $Imma$ structure are relatively small, inducing a slightly larging screening on the $d_{xz/yz}$ orbitals. For the AFM DFT+\ucrpa{} predicted $Pnma$ structure we obtained (all values in eV):
\begin{align*}
    U_{ij}^{\sigma \bar{\sigma}} &=
\begin{pmatrix}
 3.059 & 2.351 & 2.337\\
  2.351 & 3.036 & 2.329\\
  2.337 & 2.329 & 2.997\\
\end{pmatrix} \\ 
U_{ij}^{\sigma \sigma} &=
\begin{pmatrix}
  0. & 2.023 & 2.01\\
  2.023 & 0. & 2.003\\
  2.01 & 2.003 & 0.\\
\end{pmatrix}
\end{align*}
This demonstrates the effect of the additional distortions in $Pnma$, which reduces the onsite Coulomb interaction value for $d_{xy}$ compared to $Imma$. Finally, for the \pddmodel{} model in the cubic structure we obtained (all values in eV):
\begin{align*}
    U_{ij}^{\sigma \bar{\sigma}} &=
\begin{pmatrix}
  3.286 & 2.286 & 2.286 & 2.015 & 1.921\\
  2.286 & 3.017 & 1.965 & 2.043 & 1.965\\
  2.286 & 1.965 & 3.017 & 2.043 & 1.965\\
  2.015 & 2.043 & 2.043 & 3.286 & 2.408\\
  1.921 & 1.965 & 1.965 & 2.408 & 3.017\\
\end{pmatrix} \\
U_{ij}^{\sigma \sigma} &=
\begin{pmatrix}
  0. & 1.874 & 1.874 & 1.379 & 1.309\\
  1.874 & 0. & 1.451 & 1.497 & 1.451\\
  1.874 & 1.451 & 0. & 1.497 & 1.451\\
  1.379 & 1.497 & 1.497 & 0. & 2.063\\
  1.309 & 1.451 & 1.451 & 2.063 & 0.\\
\end{pmatrix}
\end{align*}

Table~\ref{tab:cRPA-comp} shows a list of the fitted cRPA results, including a comparison between the \cubic{} and the $Imma$ structure. Fig.~\ref{fig:crpa_r4} shows the on-site Coulomb interaction tensor values as function of the \rfour{} amplitude in the \ttgmodel{} model.

\begin{figure}[t]
    \centering
    \includegraphics[width=0.9\linewidth]{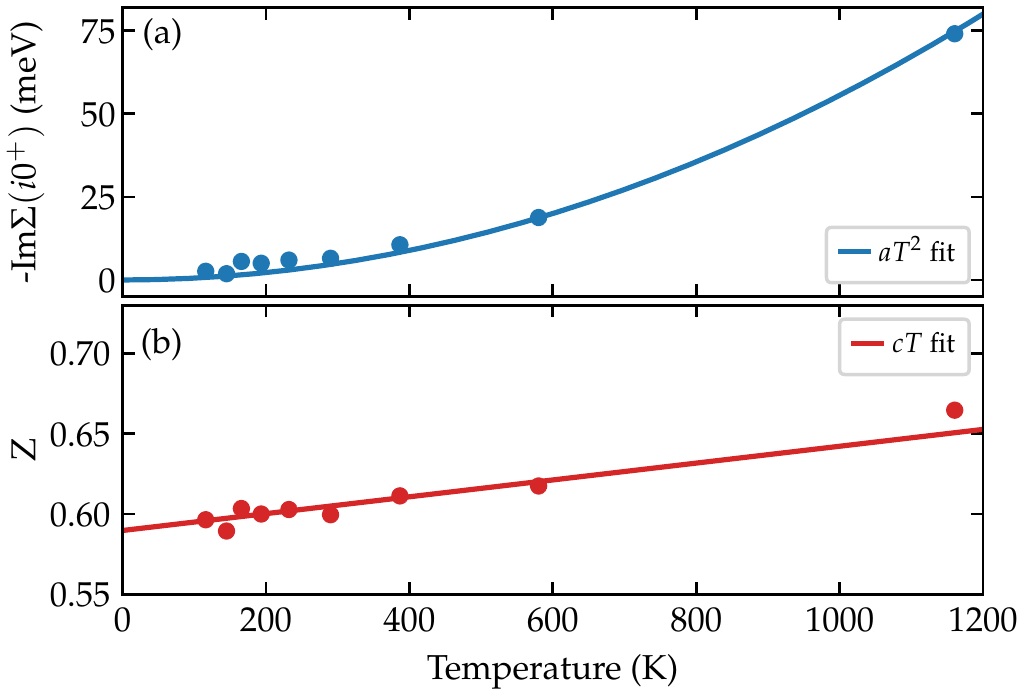}
    \caption{Comparison of (a) Quasiparticle scattering rate -Im$\Sigma(i0^+)$ and (b) quasiparticle weight $Z$ as function of temperature for the \ttgmodel{} model using cRPA interaction values for the experimental $Imma$ structure. (a) shows clear $T^2$ behavior ($a=$\SI{5.56e-5}{meV/K^2}), whereas (b) shows a modest linear temperature dependence ($c=$\SI{5.25e-5}{K{^-1}}). 
    }
    \label{fig:Z_tau_temp}
\end{figure}

\section{Quasiparticles within DMFT: physical properties and transport}
\label{app:quasiparticles}

In Fig.~\ref{fig:self_energies} the imaginary part of the DMFT $d$-orbital self-energies on the Matsubara axis are shown for various \rfour{} amplitudes, both for the \ttgmodel{} and the \pddmodel{} model. A fit to a 4th order polynomial 
over the first Matsubara frequencies yields a determination of the two key quantities $Z$ and $\text{Im}\Sigma(i 0^+)$ 
involved in the low-frequency expansion:
\begin{align}
\mathrm{Im}\Sigma(i\omega)\,=\, \mathrm{Im}\Sigma(i0^+)\,+\,i\omega\,(1-\frac{1}{Z})+\cdots   
\end{align}
with $1/Z =  1 -\partial\text{Im}\Sigma(i\omega_n)/\partial \omega_n\Bigr|_{i \omega_n \rightarrow 0}$. 
Inserting this into the expression of the Green's function: $G^{-1}=i\omega+\mu-H(\mathbf{k})-\Sigma$, expanding at low frequency, 
and focusing on the partially filled $t_{2g}$ states, 
one sees that the low-energy quasiparticles carry a spectral weight $Z$, and are characterized 
by an effective mass enhancement $m^*/m_b$ (with $m_b$ the band mass)  
and quasiparticle lifetime $\tau^*$ given by:
\begin{align}
    \frac{m^*}{m_b}\,=\,\frac{1}{Z}\,\,\,,\,\,\,\frac{1}{\tau^*}\,=\,-Z\mathrm{Im}\Sigma(i0^+)
\end{align}
The conductivity can be calculated from linear response theory with the Kubo formula. Because the self-energy is spatially local in the DMFT approximation, vertex corrections vanish and hence the transport lifetime can be directly related to single-particle 
quantities (see e.g., Refs.~\onlinecite{Georges:1996,Berthod:2013,Xiaoyu2014}). Specifically, one obtains for the direct current (dc) conductivity (quoting the formula for a single band for simplicity)
\begin{align}
\sigma_{dc}\,=\,\omega_{P0}^2\,\tau_{\mathrm{tr}}
\end{align}
in which $\omega_{P0}$ is the plasma frequency obtained within band-structure theory (i.e. unrenormalized by correlations) 
and $\tau_{\mathrm{tr}}^{-1}=-2\mathrm{Im}\Sigma(i0^+)$ \cite{Berthod:2013}. Note that, importantly, the quasiparticle weight $Z$ does not enter this expression and drops out from transport properties. This can also be understood from a Boltzmann transport description 
in terms of quasiparticles~\cite{Xiaoyu2014}. 
In that view, the plasma frequency is the renormalized one involving the quasiparticle effective 
mass, and thus is multiplied by $Z$ as compared to the bare one, and the lifetime is related to the quasiparticle lifetime $\tau^*$ given above. As a result, $Z$ drops out form the product and we recover the expression above for $\sigma_{dc}$. 

In Fig.~\ref{fig:Z_tau_temp}(a) the temperature dependence of $-\mathrm{Im}\Sigma(i0^+)$ is shown, which is found to 
be consistent with the $T^2$ Fermi liquid behavior observed experimentally~\cite{Nagai2005}. Remarkably, the quadratic behavior found in our calculations of $-\mathrm{Im}\Sigma(i0^+)$ extends to very high temperatures up to 1200~K, indicated by a $aT^2$ fit in Fig.~\ref{fig:Z_tau_temp}(a). At temperatures below 200~K small deviations from the fit can be observed, which are probably due to uncertainties in the calculation of $-\mathrm{Im}\Sigma(i0^+)$. Calculating the scattering rate by fitting a polynomial to the lowest Matsubara frequencies at different temperatures, and the numerical noise at lower temperatures, complicates the evaluation of the scattering rate. Overall, we find good agreement with our fit, indicating correctly the trend to a vanishing $-\mathrm{Im}\Sigma(i0^+)$ at 0~K for the inelastic scattering described by DMFT.

\bibliography{bibfile}

\begin{thebibliography}{64}%
\makeatletter
\providecommand \@ifxundefined [1]{%
 \@ifx{#1\undefined}
}%
\providecommand \@ifnum [1]{%
 \ifnum #1\expandafter \@firstoftwo
 \else \expandafter \@secondoftwo
 \fi
}%
\providecommand \@ifx [1]{%
 \ifx #1\expandafter \@firstoftwo
 \else \expandafter \@secondoftwo
 \fi
}%
\providecommand \natexlab [1]{#1}%
\providecommand \enquote  [1]{``#1''}%
\providecommand \bibnamefont  [1]{#1}%
\providecommand \bibfnamefont [1]{#1}%
\providecommand \citenamefont [1]{#1}%
\providecommand \href@noop [0]{\@secondoftwo}%
\providecommand \href [0]{\begingroup \@sanitize@url \@href}%
\providecommand \@href[1]{\@@startlink{#1}\@@href}%
\providecommand \@@href[1]{\endgroup#1\@@endlink}%
\providecommand \@sanitize@url [0]{\catcode `\\12\catcode `\$12\catcode
  `\&12\catcode `\#12\catcode `\^12\catcode `\_12\catcode `\%12\relax}%
\providecommand \@@startlink[1]{}%
\providecommand \@@endlink[0]{}%
\providecommand \url  [0]{\begingroup\@sanitize@url \@url }%
\providecommand \@url [1]{\endgroup\@href {#1}{\urlprefix }}%
\providecommand \urlprefix  [0]{URL }%
\providecommand \Eprint [0]{\href }%
\providecommand \doibase [0]{https://doi.org/}%
\providecommand \selectlanguage [0]{\@gobble}%
\providecommand \bibinfo  [0]{\@secondoftwo}%
\providecommand \bibfield  [0]{\@secondoftwo}%
\providecommand \translation [1]{[#1]}%
\providecommand \BibitemOpen [0]{}%
\providecommand \bibitemStop [0]{}%
\providecommand \bibitemNoStop [0]{.\EOS\space}%
\providecommand \EOS [0]{\spacefactor3000\relax}%
\providecommand \BibitemShut  [1]{\csname bibitem#1\endcsname}%
\let\auto@bib@innerbib\@empty
\bibitem [{\citenamefont {Dagotto}(1994)}]{RevModPhys.66.763}%
  \BibitemOpen
  \bibfield  {author} {\bibinfo {author} {\bibfnamefont {E.}~\bibnamefont
  {Dagotto}},\ }\href {https://doi.org/10.1103/RevModPhys.66.763} {\bibfield
  {journal} {\bibinfo  {journal} {Reviews of Modern Physics}\ }\textbf
  {\bibinfo {volume} {66}},\ \bibinfo {pages} {763} (\bibinfo {year}
  {1994})}\BibitemShut {NoStop}%
\bibitem [{\citenamefont {Stewart}(2001)}]{RevModPhys.73.797}%
  \BibitemOpen
  \bibfield  {author} {\bibinfo {author} {\bibfnamefont {G.~R.}\ \bibnamefont
  {Stewart}},\ }\href {https://doi.org/10.1103/RevModPhys.73.797} {\bibfield
  {journal} {\bibinfo  {journal} {Reviews of Modern Physics}\ }\textbf
  {\bibinfo {volume} {73}},\ \bibinfo {pages} {797} (\bibinfo {year}
  {2001})}\BibitemShut {NoStop}%
\bibitem [{\citenamefont {Khomskii}(2009)}]{Khomskii2009}%
  \BibitemOpen
  \bibfield  {author} {\bibinfo {author} {\bibfnamefont {D.}~\bibnamefont
  {Khomskii}},\ }\href@noop {} {\bibfield  {journal} {\bibinfo  {journal}
  {Physics}\ }\textbf {\bibinfo {volume} {2}},\ \bibinfo {pages} {20} (\bibinfo
  {year} {2009})}\BibitemShut {NoStop}%
\bibitem [{\citenamefont {Haule}(2018)}]{Haule2018}%
  \BibitemOpen
  \bibfield  {author} {\bibinfo {author} {\bibfnamefont {K.}~\bibnamefont
  {Haule}},\ }\href {https://doi.org/10.7566/JPSJ.87.041005} {\bibfield
  {journal} {\bibinfo  {journal} {Journal of the Physical Society of Japan}\
  }\textbf {\bibinfo {volume} {87}},\ \bibinfo {pages} {041005} (\bibinfo
  {year} {2018})}\BibitemShut {NoStop}%
\bibitem [{\citenamefont {Imada}\ \emph {et~al.}(1998)\citenamefont {Imada},
  \citenamefont {Fujimori},\ and\ \citenamefont {Tokura}}]{Imada1998}%
  \BibitemOpen
  \bibfield  {author} {\bibinfo {author} {\bibfnamefont {M.}~\bibnamefont
  {Imada}}, \bibinfo {author} {\bibfnamefont {A.}~\bibnamefont {Fujimori}},\
  and\ \bibinfo {author} {\bibfnamefont {Y.}~\bibnamefont {Tokura}},\ }\href
  {https://doi.org/10.1103/RevModPhys.70.1039} {\bibfield  {journal} {\bibinfo
  {journal} {Reviews of Modern Physics}\ }\textbf {\bibinfo {volume} {70}},\
  \bibinfo {pages} {1039} (\bibinfo {year} {1998})}\BibitemShut {NoStop}%
\bibitem [{\citenamefont {Guzm{\'a}n-Verri}\ \emph {et~al.}(2019)\citenamefont
  {Guzm{\'a}n-Verri}, \citenamefont {Brierley},\ and\ \citenamefont
  {Littlewood}}]{Guzman-Verri2019}%
  \BibitemOpen
  \bibfield  {author} {\bibinfo {author} {\bibfnamefont {G.~G.}\ \bibnamefont
  {Guzm{\'a}n-Verri}}, \bibinfo {author} {\bibfnamefont {R.~T.}\ \bibnamefont
  {Brierley}},\ and\ \bibinfo {author} {\bibfnamefont {P.~B.}\ \bibnamefont
  {Littlewood}},\ }\href {https://doi.org/10.1038/s41586-019-1824-9} {\bibfield
   {journal} {\bibinfo  {journal} {Nature}\ }\textbf {\bibinfo {volume}
  {576}},\ \bibinfo {pages} {429} (\bibinfo {year} {2019})}\BibitemShut
  {NoStop}%
\bibitem [{\citenamefont {Cammarata}\ and\ \citenamefont
  {Rondinelli}(2014)}]{Cammarata2014}%
  \BibitemOpen
  \bibfield  {author} {\bibinfo {author} {\bibfnamefont {A.}~\bibnamefont
  {Cammarata}}\ and\ \bibinfo {author} {\bibfnamefont {J.~M.}\ \bibnamefont
  {Rondinelli}},\ }\href {https://doi.org/10.1063/1.4895967} {\bibfield
  {journal} {\bibinfo  {journal} {The Journal of Chemical Physics}\ }\textbf
  {\bibinfo {volume} {141}},\ \bibinfo {pages} {114704} (\bibinfo {year}
  {2014})}\BibitemShut {NoStop}%
\bibitem [{\citenamefont {Leonov}\ \emph {et~al.}(2008)\citenamefont {Leonov},
  \citenamefont {Binggeli}, \citenamefont {Korotin}, \citenamefont {Anisimov},
  \citenamefont {Stoji\ifmmode~\acute{c}\else \'{c}\fi{}},\ and\ \citenamefont
  {Vollhardt}}]{leonov2008}%
  \BibitemOpen
  \bibfield  {author} {\bibinfo {author} {\bibfnamefont {I.}~\bibnamefont
  {Leonov}}, \bibinfo {author} {\bibfnamefont {N.}~\bibnamefont {Binggeli}},
  \bibinfo {author} {\bibfnamefont {D.}~\bibnamefont {Korotin}}, \bibinfo
  {author} {\bibfnamefont {V.~I.}\ \bibnamefont {Anisimov}}, \bibinfo {author}
  {\bibfnamefont {N.}~\bibnamefont {Stoji\ifmmode~\acute{c}\else \'{c}\fi{}}},\
  and\ \bibinfo {author} {\bibfnamefont {D.}~\bibnamefont {Vollhardt}},\ }\href
  {https://doi.org/10.1103/PhysRevLett.101.096405} {\bibfield  {journal}
  {\bibinfo  {journal} {Physical Review Letters}\ }\textbf {\bibinfo {volume}
  {101}},\ \bibinfo {pages} {096405} (\bibinfo {year} {2008})}\BibitemShut
  {NoStop}%
\bibitem [{\citenamefont {Peil}\ \emph {et~al.}(2019)\citenamefont {Peil},
  \citenamefont {Hampel}, \citenamefont {Ederer},\ and\ \citenamefont
  {Georges}}]{peil:2019}%
  \BibitemOpen
  \bibfield  {author} {\bibinfo {author} {\bibfnamefont {O.~E.}\ \bibnamefont
  {Peil}}, \bibinfo {author} {\bibfnamefont {A.}~\bibnamefont {Hampel}},
  \bibinfo {author} {\bibfnamefont {C.}~\bibnamefont {Ederer}},\ and\ \bibinfo
  {author} {\bibfnamefont {A.}~\bibnamefont {Georges}},\ }\href
  {https://doi.org/10.1103/PhysRevB.99.245127} {\bibfield  {journal} {\bibinfo
  {journal} {Physical Review B}\ }\textbf {\bibinfo {volume} {99}},\ \bibinfo
  {pages} {245127} (\bibinfo {year} {2019})}\BibitemShut {NoStop}%
\bibitem [{\citenamefont {Park}\ \emph {et~al.}(2014)\citenamefont {Park},
  \citenamefont {Millis},\ and\ \citenamefont {Marianetti}}]{Park2014short}%
  \BibitemOpen
  \bibfield  {author} {\bibinfo {author} {\bibfnamefont {H.}~\bibnamefont
  {Park}}, \bibinfo {author} {\bibfnamefont {A.~J.}\ \bibnamefont {Millis}},\
  and\ \bibinfo {author} {\bibfnamefont {C.~A.}\ \bibnamefont {Marianetti}},\
  }\href {https://doi.org/10.1103/PhysRevB.89.245133} {\bibfield  {journal}
  {\bibinfo  {journal} {Physical Review B}\ }\textbf {\bibinfo {volume} {89}},\
  \bibinfo {pages} {245133} (\bibinfo {year} {2014})}\BibitemShut {NoStop}%
\bibitem [{\citenamefont {Balachandran}\ \emph {et~al.}(2018)\citenamefont
  {Balachandran}, \citenamefont {Emery}, \citenamefont {Gubernatis},
  \citenamefont {Lookman}, \citenamefont {Wolverton},\ and\ \citenamefont
  {Zunger}}]{Balachandran2018}%
  \BibitemOpen
  \bibfield  {author} {\bibinfo {author} {\bibfnamefont {P.~V.}\ \bibnamefont
  {Balachandran}}, \bibinfo {author} {\bibfnamefont {A.~A.}\ \bibnamefont
  {Emery}}, \bibinfo {author} {\bibfnamefont {J.~E.}\ \bibnamefont
  {Gubernatis}}, \bibinfo {author} {\bibfnamefont {T.}~\bibnamefont {Lookman}},
  \bibinfo {author} {\bibfnamefont {C.}~\bibnamefont {Wolverton}},\ and\
  \bibinfo {author} {\bibfnamefont {A.}~\bibnamefont {Zunger}},\ }\href
  {https://doi.org/10.1103/PhysRevMaterials.2.043802} {\bibfield  {journal}
  {\bibinfo  {journal} {Phys. Rev. Materials}\ }\textbf {\bibinfo {volume}
  {2}},\ \bibinfo {pages} {043802} (\bibinfo {year} {2018})}\BibitemShut
  {NoStop}%
\bibitem [{\citenamefont {Hong}\ \emph {et~al.}(2012)\citenamefont {Hong},
  \citenamefont {Stroppa}, \citenamefont {\'I\~niguez}, \citenamefont
  {Picozzi},\ and\ \citenamefont {Vanderbilt}}]{PhysRevB.85.054417}%
  \BibitemOpen
  \bibfield  {author} {\bibinfo {author} {\bibfnamefont {J.}~\bibnamefont
  {Hong}}, \bibinfo {author} {\bibfnamefont {A.}~\bibnamefont {Stroppa}},
  \bibinfo {author} {\bibfnamefont {J.}~\bibnamefont {\'I\~niguez}}, \bibinfo
  {author} {\bibfnamefont {S.}~\bibnamefont {Picozzi}},\ and\ \bibinfo {author}
  {\bibfnamefont {D.}~\bibnamefont {Vanderbilt}},\ }\href
  {https://doi.org/10.1103/PhysRevB.85.054417} {\bibfield  {journal} {\bibinfo
  {journal} {Physical Review B}\ }\textbf {\bibinfo {volume} {85}},\ \bibinfo
  {pages} {054417} (\bibinfo {year} {2012})}\BibitemShut {NoStop}%
\bibitem [{\citenamefont {Varignon}\ \emph {et~al.}(2019)\citenamefont
  {Varignon}, \citenamefont {Bibes},\ and\ \citenamefont
  {Zunger}}]{Varignon2019}%
  \BibitemOpen
  \bibfield  {author} {\bibinfo {author} {\bibfnamefont {J.}~\bibnamefont
  {Varignon}}, \bibinfo {author} {\bibfnamefont {M.}~\bibnamefont {Bibes}},\
  and\ \bibinfo {author} {\bibfnamefont {A.}~\bibnamefont {Zunger}},\ }\href
  {https://doi.org/10.1038/s41467-019-09698-6} {\bibfield  {journal} {\bibinfo
  {journal} {Nature Communications}\ }\textbf {\bibinfo {volume} {10}},\
  \bibinfo {pages} {1658} (\bibinfo {year} {2019})}\BibitemShut {NoStop}%
\bibitem [{\citenamefont {Ramberger}\ \emph {et~al.}(2017)\citenamefont
  {Ramberger}, \citenamefont {Sch\"afer},\ and\ \citenamefont
  {Kresse}}]{Ramberger2017}%
  \BibitemOpen
  \bibfield  {author} {\bibinfo {author} {\bibfnamefont {B.}~\bibnamefont
  {Ramberger}}, \bibinfo {author} {\bibfnamefont {T.}~\bibnamefont
  {Sch\"afer}},\ and\ \bibinfo {author} {\bibfnamefont {G.}~\bibnamefont
  {Kresse}},\ }\href {https://doi.org/10.1103/PhysRevLett.118.106403}
  {\bibfield  {journal} {\bibinfo  {journal} {Physical Review Letters}\
  }\textbf {\bibinfo {volume} {118}},\ \bibinfo {pages} {106403} (\bibinfo
  {year} {2017})}\BibitemShut {NoStop}%
\bibitem [{\citenamefont {Kotliar}\ \emph {et~al.}(2006)\citenamefont
  {Kotliar}, \citenamefont {Savrasov}, \citenamefont {Haule}, \citenamefont
  {Oudovenko}, \citenamefont {Parcollet},\ and\ \citenamefont
  {Marianetti}}]{Kotliar:2006}%
  \BibitemOpen
  \bibfield  {author} {\bibinfo {author} {\bibfnamefont {G.}~\bibnamefont
  {Kotliar}}, \bibinfo {author} {\bibfnamefont {S.~Y.}\ \bibnamefont
  {Savrasov}}, \bibinfo {author} {\bibfnamefont {K.}~\bibnamefont {Haule}},
  \bibinfo {author} {\bibfnamefont {V.~S.}\ \bibnamefont {Oudovenko}}, \bibinfo
  {author} {\bibfnamefont {O.}~\bibnamefont {Parcollet}},\ and\ \bibinfo
  {author} {\bibfnamefont {C.~A.}\ \bibnamefont {Marianetti}},\ }\href
  {https://doi.org/10.1103/RevModPhys.78.865} {\bibfield  {journal} {\bibinfo
  {journal} {Reviews of Modern Physics}\ }\textbf {\bibinfo {volume} {78}},\
  \bibinfo {pages} {865} (\bibinfo {year} {2006})}\BibitemShut {NoStop}%
\bibitem [{\citenamefont {Held}(2007)}]{held2007}%
  \BibitemOpen
  \bibfield  {author} {\bibinfo {author} {\bibfnamefont {K.}~\bibnamefont
  {Held}},\ }\href {https://doi.org/10.1080/00018730701619647} {\bibfield
  {journal} {\bibinfo  {journal} {Advances in Physics}\ }\textbf {\bibinfo
  {volume} {56}},\ \bibinfo {pages} {829} (\bibinfo {year} {2007})}\BibitemShut
  {NoStop}%
\bibitem [{\citenamefont {Ikeda}\ and\ \citenamefont
  {Shirakawa}(2000)}]{Ikeda2000}%
  \BibitemOpen
  \bibfield  {author} {\bibinfo {author} {\bibfnamefont {S.~I.}\ \bibnamefont
  {Ikeda}}\ and\ \bibinfo {author} {\bibfnamefont {N.}~\bibnamefont
  {Shirakawa}},\ }\href {https://doi.org/10.1016/S0921-4534(00)00692-4}
  {\bibfield  {journal} {\bibinfo  {journal} {Physica C: Superconductivity and
  its Applications}\ }\textbf {\bibinfo {volume} {341-348}},\ \bibinfo {pages}
  {785} (\bibinfo {year} {2000})}\BibitemShut {NoStop}%
\bibitem [{\citenamefont {Macquart}\ \emph {et~al.}(2010)\citenamefont
  {Macquart}, \citenamefont {Kennedy},\ and\ \citenamefont
  {Avdeev}}]{Macquart2010}%
  \BibitemOpen
  \bibfield  {author} {\bibinfo {author} {\bibfnamefont {B.}~\bibnamefont
  {Macquart}}, \bibinfo {author} {\bibfnamefont {B.~J.}\ \bibnamefont
  {Kennedy}},\ and\ \bibinfo {author} {\bibfnamefont {M.}~\bibnamefont
  {Avdeev}},\ }\href {https://doi.org/10.1016/j.jssc.2009.11.005} {\bibfield
  {journal} {\bibinfo  {journal} {Journal of Solid State Chemistry}\ }\textbf
  {\bibinfo {volume} {183}},\ \bibinfo {pages} {249} (\bibinfo {year}
  {2010})}\BibitemShut {NoStop}%
\bibitem [{\citenamefont {Nagai}\ \emph {et~al.}(2005)\citenamefont {Nagai},
  \citenamefont {Shirakawa}, \citenamefont {Ikeda}, \citenamefont {Iwasaki},
  \citenamefont {Nishimura},\ and\ \citenamefont {Kosaka}}]{Nagai2005}%
  \BibitemOpen
  \bibfield  {author} {\bibinfo {author} {\bibfnamefont {I.}~\bibnamefont
  {Nagai}}, \bibinfo {author} {\bibfnamefont {N.}~\bibnamefont {Shirakawa}},
  \bibinfo {author} {\bibfnamefont {S.~I.}\ \bibnamefont {Ikeda}}, \bibinfo
  {author} {\bibfnamefont {R.}~\bibnamefont {Iwasaki}}, \bibinfo {author}
  {\bibfnamefont {H.}~\bibnamefont {Nishimura}},\ and\ \bibinfo {author}
  {\bibfnamefont {M.}~\bibnamefont {Kosaka}},\ }\href
  {https://doi.org/10.1063/1.1992671} {\bibfield  {journal} {\bibinfo
  {journal} {Applied Physics Letters}\ }\textbf {\bibinfo {volume} {87}},\
  \bibinfo {pages} {2} (\bibinfo {year} {2005})}\BibitemShut {NoStop}%
\bibitem [{\citenamefont {Wadati}\ \emph {et~al.}(2014)\citenamefont {Wadati},
  \citenamefont {Mravlje}, \citenamefont {Yoshimatsu}, \citenamefont
  {Kumigashira}, \citenamefont {Oshima}, \citenamefont {Sugiyama},
  \citenamefont {Ikenaga}, \citenamefont {Fujimori}, \citenamefont {Georges},
  \citenamefont {Radetinac}, \citenamefont {Takahashi}, \citenamefont
  {Kawasaki},\ and\ \citenamefont {Tokura}}]{Wadati2014}%
  \BibitemOpen
  \bibfield  {author} {\bibinfo {author} {\bibfnamefont {H.}~\bibnamefont
  {Wadati}}, \bibinfo {author} {\bibfnamefont {J.}~\bibnamefont {Mravlje}},
  \bibinfo {author} {\bibfnamefont {K.}~\bibnamefont {Yoshimatsu}}, \bibinfo
  {author} {\bibfnamefont {H.}~\bibnamefont {Kumigashira}}, \bibinfo {author}
  {\bibfnamefont {M.}~\bibnamefont {Oshima}}, \bibinfo {author} {\bibfnamefont
  {T.}~\bibnamefont {Sugiyama}}, \bibinfo {author} {\bibfnamefont
  {E.}~\bibnamefont {Ikenaga}}, \bibinfo {author} {\bibfnamefont
  {A.}~\bibnamefont {Fujimori}}, \bibinfo {author} {\bibfnamefont
  {A.}~\bibnamefont {Georges}}, \bibinfo {author} {\bibfnamefont
  {A.}~\bibnamefont {Radetinac}}, \bibinfo {author} {\bibfnamefont {K.~S.}\
  \bibnamefont {Takahashi}}, \bibinfo {author} {\bibfnamefont {M.}~\bibnamefont
  {Kawasaki}},\ and\ \bibinfo {author} {\bibfnamefont {Y.}~\bibnamefont
  {Tokura}},\ }\href {https://doi.org/10.1103/PhysRevB.90.205131} {\bibfield
  {journal} {\bibinfo  {journal} {Physical Review B}\ }\textbf {\bibinfo
  {volume} {90}},\ \bibinfo {pages} {205131} (\bibinfo {year}
  {2014})}\BibitemShut {NoStop}%
\bibitem [{\citenamefont {Somia}\ \emph {et~al.}(2019)\citenamefont {Somia},
  \citenamefont {Mehmood}, \citenamefont {Ali}, \citenamefont {Khan},
  \citenamefont {Khan},\ and\ \citenamefont {Ahmad}}]{Somia2019}%
  \BibitemOpen
  \bibfield  {author} {\bibinfo {author} {\bibnamefont {Somia}}, \bibinfo
  {author} {\bibfnamefont {S.}~\bibnamefont {Mehmood}}, \bibinfo {author}
  {\bibfnamefont {Z.}~\bibnamefont {Ali}}, \bibinfo {author} {\bibfnamefont
  {I.}~\bibnamefont {Khan}}, \bibinfo {author} {\bibfnamefont {F.}~\bibnamefont
  {Khan}},\ and\ \bibinfo {author} {\bibfnamefont {I.~a.}\ \bibnamefont
  {Ahmad}},\ }\href {https://doi.org/10.1007/s11664-018-06870-4} {\bibfield
  {journal} {\bibinfo  {journal} {Journal of Electronic Materials}\ }\textbf
  {\bibinfo {volume} {48}},\ \bibinfo {pages} {1730} (\bibinfo {year}
  {2019})}\BibitemShut {NoStop}%
\bibitem [{\citenamefont {Zhu}\ \emph {et~al.}(2012)\citenamefont {Zhu},
  \citenamefont {Gu}, \citenamefont {Jia},\ and\ \citenamefont {Hu}}]{Zhu2012}%
  \BibitemOpen
  \bibfield  {author} {\bibinfo {author} {\bibfnamefont {Z.~L.}\ \bibnamefont
  {Zhu}}, \bibinfo {author} {\bibfnamefont {J.~H.}\ \bibnamefont {Gu}},
  \bibinfo {author} {\bibfnamefont {Y.}~\bibnamefont {Jia}},\ and\ \bibinfo
  {author} {\bibfnamefont {X.}~\bibnamefont {Hu}},\ }\href
  {https://doi.org/10.1016/j.physb.2012.01.126} {\bibfield  {journal} {\bibinfo
   {journal} {Physica B: Condensed Matter}\ }\textbf {\bibinfo {volume}
  {407}},\ \bibinfo {pages} {1990} (\bibinfo {year} {2012})}\BibitemShut
  {NoStop}%
\bibitem [{\citenamefont {Tariq}\ \emph {et~al.}(2018)\citenamefont {Tariq},
  \citenamefont {Jamil}, \citenamefont {Sharif}, \citenamefont {Ramay},
  \citenamefont {Ahmad}, \citenamefont {ul~Qamar},\ and\ \citenamefont
  {Tahir}}]{Tariq2018}%
  \BibitemOpen
  \bibfield  {author} {\bibinfo {author} {\bibfnamefont {S.}~\bibnamefont
  {Tariq}}, \bibinfo {author} {\bibfnamefont {M.~I.}\ \bibnamefont {Jamil}},
  \bibinfo {author} {\bibfnamefont {A.}~\bibnamefont {Sharif}}, \bibinfo
  {author} {\bibfnamefont {S.~M.}\ \bibnamefont {Ramay}}, \bibinfo {author}
  {\bibfnamefont {H.~n.}\ \bibnamefont {Ahmad}}, \bibinfo {author}
  {\bibfnamefont {N.}~\bibnamefont {ul~Qamar}},\ and\ \bibinfo {author}
  {\bibfnamefont {B.}~\bibnamefont {Tahir}},\ }\href
  {https://doi.org/10.1007/s00339-017-1452-x} {\bibfield  {journal} {\bibinfo
  {journal} {Applied Physics A: Materials Science and Processing}\ }\textbf
  {\bibinfo {volume} {124}},\ \bibinfo {pages} {1} (\bibinfo {year}
  {2018})}\BibitemShut {NoStop}%
\bibitem [{\citenamefont {Lee-Hand}\ \emph {et~al.}(2020)\citenamefont
  {Lee-Hand}, \citenamefont {Hampel},\ and\ \citenamefont
  {Dreyer}}]{LeeHand2020}%
  \BibitemOpen
  \bibfield  {author} {\bibinfo {author} {\bibfnamefont {J.}~\bibnamefont
  {Lee-Hand}}, \bibinfo {author} {\bibfnamefont {A.}~\bibnamefont {Hampel}},\
  and\ \bibinfo {author} {\bibfnamefont {C.~E.}\ \bibnamefont {Dreyer}},\
  }\href@noop {} {} (\bibinfo {year} {2020}),\ \Eprint
  {https://arxiv.org/abs/2011.08323} {arXiv:2011.08323} \BibitemShut {NoStop}%
\bibitem [{\citenamefont {Perez-Mato}\ \emph {et~al.}(2010)\citenamefont
  {Perez-Mato}, \citenamefont {Orobengoa},\ and\ \citenamefont
  {Aroyo}}]{PerezMato:2010ix}%
  \BibitemOpen
  \bibfield  {author} {\bibinfo {author} {\bibfnamefont {J.~M.}\ \bibnamefont
  {Perez-Mato}}, \bibinfo {author} {\bibfnamefont {D.}~\bibnamefont
  {Orobengoa}},\ and\ \bibinfo {author} {\bibfnamefont {M.~I.}\ \bibnamefont
  {Aroyo}},\ }\href@noop {} {\bibfield  {journal} {\bibinfo  {journal} {Acta
  Crystallographica A}\ }\textbf {\bibinfo {volume} {66}},\ \bibinfo {pages}
  {558} (\bibinfo {year} {2010})}\BibitemShut {NoStop}%
\bibitem [{\citenamefont {Campbell}\ \emph {et~al.}(2006)\citenamefont
  {Campbell}, \citenamefont {Stokes}, \citenamefont {Tanner},\ and\
  \citenamefont {Hatch}}]{Campbell:2006}%
  \BibitemOpen
  \bibfield  {author} {\bibinfo {author} {\bibfnamefont {B.~J.}\ \bibnamefont
  {Campbell}}, \bibinfo {author} {\bibfnamefont {H.~T.}\ \bibnamefont
  {Stokes}}, \bibinfo {author} {\bibfnamefont {D.~E.}\ \bibnamefont {Tanner}},\
  and\ \bibinfo {author} {\bibfnamefont {D.~M.}\ \bibnamefont {Hatch}},\
  }\href@noop {} {\bibfield  {journal} {\bibinfo  {journal} {Journal of Applied
  Crystallography}\ }\textbf {\bibinfo {volume} {39}},\ \bibinfo {pages} {607}
  (\bibinfo {year} {2006})}\BibitemShut {NoStop}%
\bibitem [{\citenamefont {Bl{\"o}chl}(1994)}]{Blochl:1994dx}%
  \BibitemOpen
  \bibfield  {author} {\bibinfo {author} {\bibfnamefont {P.~E.}\ \bibnamefont
  {Bl{\"o}chl}},\ }\href@noop {} {\bibfield  {journal} {\bibinfo  {journal}
  {Physical Review B}\ }\textbf {\bibinfo {volume} {50}},\ \bibinfo {pages}
  {17953} (\bibinfo {year} {1994})}\BibitemShut {NoStop}%
\bibitem [{\citenamefont {Kresse}\ and\ \citenamefont
  {Hafner}(1993)}]{Kresse:1993bz}%
  \BibitemOpen
  \bibfield  {author} {\bibinfo {author} {\bibfnamefont {G.}~\bibnamefont
  {Kresse}}\ and\ \bibinfo {author} {\bibfnamefont {J.}~\bibnamefont
  {Hafner}},\ }\href {https://doi.org/10.1103/PhysRevB.47.558} {\bibfield
  {journal} {\bibinfo  {journal} {Physical Review B}\ }\textbf {\bibinfo
  {volume} {47}},\ \bibinfo {pages} {558} (\bibinfo {year} {1993})}\BibitemShut
  {NoStop}%
\bibitem [{\citenamefont {Kresse}\ and\ \citenamefont
  {Furthm\"uller}(1996)}]{Kresse:1996kl}%
  \BibitemOpen
  \bibfield  {author} {\bibinfo {author} {\bibfnamefont {G.}~\bibnamefont
  {Kresse}}\ and\ \bibinfo {author} {\bibfnamefont {J.}~\bibnamefont
  {Furthm\"uller}},\ }\href {https://doi.org/10.1103/PhysRevB.54.11169}
  {\bibfield  {journal} {\bibinfo  {journal} {Physical Review B}\ }\textbf
  {\bibinfo {volume} {54}},\ \bibinfo {pages} {11169} (\bibinfo {year}
  {1996})}\BibitemShut {NoStop}%
\bibitem [{\citenamefont {Kresse}\ and\ \citenamefont
  {Joubert}(1999)}]{Kresse:1999dk}%
  \BibitemOpen
  \bibfield  {author} {\bibinfo {author} {\bibfnamefont {G.}~\bibnamefont
  {Kresse}}\ and\ \bibinfo {author} {\bibfnamefont {D.}~\bibnamefont
  {Joubert}},\ }\href {https://doi.org/10.1103/PhysRevB.59.1758} {\bibfield
  {journal} {\bibinfo  {journal} {Physical Review B}\ }\textbf {\bibinfo
  {volume} {59}},\ \bibinfo {pages} {1758} (\bibinfo {year}
  {1999})}\BibitemShut {NoStop}%
\bibitem [{\citenamefont {Perdew}\ \emph {et~al.}(1996)\citenamefont {Perdew},
  \citenamefont {Burke},\ and\ \citenamefont {Ernzerhof}}]{Perdew:1996iq}%
  \BibitemOpen
  \bibfield  {author} {\bibinfo {author} {\bibfnamefont {J.~P.}\ \bibnamefont
  {Perdew}}, \bibinfo {author} {\bibfnamefont {K.}~\bibnamefont {Burke}},\ and\
  \bibinfo {author} {\bibfnamefont {M.}~\bibnamefont {Ernzerhof}},\ }\href
  {https://doi.org/10.1103/PhysRevLett.77.3865} {\bibfield  {journal} {\bibinfo
   {journal} {Physical Review Letters}\ }\textbf {\bibinfo {volume} {77}},\
  \bibinfo {pages} {3865} (\bibinfo {year} {1996})}\BibitemShut {NoStop}%
\bibitem [{\citenamefont {Liechtenstein}\ \emph {et~al.}(1995)\citenamefont
  {Liechtenstein}, \citenamefont {Anisimov},\ and\ \citenamefont
  {Zaanen}}]{Liechtenstein:1995ip}%
  \BibitemOpen
  \bibfield  {author} {\bibinfo {author} {\bibfnamefont {A.~I.}\ \bibnamefont
  {Liechtenstein}}, \bibinfo {author} {\bibfnamefont {V.~I.}\ \bibnamefont
  {Anisimov}},\ and\ \bibinfo {author} {\bibfnamefont {J.}~\bibnamefont
  {Zaanen}},\ }\href@noop {} {\bibfield  {journal} {\bibinfo  {journal}
  {Physical Review B}\ }\textbf {\bibinfo {volume} {52}},\ \bibinfo {pages}
  {R5467} (\bibinfo {year} {1995})}\BibitemShut {NoStop}%
\bibitem [{\citenamefont {Togo}\ and\ \citenamefont {Tanaka}(2015)}]{phonopy}%
  \BibitemOpen
  \bibfield  {author} {\bibinfo {author} {\bibfnamefont {A.}~\bibnamefont
  {Togo}}\ and\ \bibinfo {author} {\bibfnamefont {I.}~\bibnamefont {Tanaka}},\
  }\href@noop {} {\bibfield  {journal} {\bibinfo  {journal} {Scr. Mater.}\
  }\textbf {\bibinfo {volume} {108}},\ \bibinfo {pages} {1} (\bibinfo {year}
  {2015})}\BibitemShut {NoStop}%
\bibitem [{\citenamefont {Amadon}\ \emph {et~al.}(2008)\citenamefont {Amadon},
  \citenamefont {Lechermann}, \citenamefont {Georges}, \citenamefont {Jollet},
  \citenamefont {Wehling},\ and\ \citenamefont {Lichtenstein}}]{Amadon:2008}%
  \BibitemOpen
  \bibfield  {author} {\bibinfo {author} {\bibfnamefont {B.}~\bibnamefont
  {Amadon}}, \bibinfo {author} {\bibfnamefont {F.}~\bibnamefont {Lechermann}},
  \bibinfo {author} {\bibfnamefont {A.}~\bibnamefont {Georges}}, \bibinfo
  {author} {\bibfnamefont {F.}~\bibnamefont {Jollet}}, \bibinfo {author}
  {\bibfnamefont {T.~O.}\ \bibnamefont {Wehling}},\ and\ \bibinfo {author}
  {\bibfnamefont {A.~I.}\ \bibnamefont {Lichtenstein}},\ }\href
  {https://doi.org/10.1103/PhysRevB.77.205112} {\bibfield  {journal} {\bibinfo
  {journal} {Physical Review B}\ }\textbf {\bibinfo {volume} {77}},\ \bibinfo
  {pages} {205112} (\bibinfo {year} {2008})}\BibitemShut {NoStop}%
\bibitem [{\citenamefont {Sch{\"u}ler}\ \emph {et~al.}(2018)\citenamefont
  {Sch{\"u}ler}, \citenamefont {Peil}, \citenamefont {Kraberger}, \citenamefont
  {Pordzik}, \citenamefont {Marsman}, \citenamefont {Kresse}, \citenamefont
  {Wehling},\ and\ \citenamefont {Aichhorn}}]{Schuler_Aichhorn:2018}%
  \BibitemOpen
  \bibfield  {author} {\bibinfo {author} {\bibfnamefont {M.}~\bibnamefont
  {Sch{\"u}ler}}, \bibinfo {author} {\bibfnamefont {O.~E.}\ \bibnamefont
  {Peil}}, \bibinfo {author} {\bibfnamefont {G.~J.}\ \bibnamefont {Kraberger}},
  \bibinfo {author} {\bibfnamefont {R.}~\bibnamefont {Pordzik}}, \bibinfo
  {author} {\bibfnamefont {M.}~\bibnamefont {Marsman}}, \bibinfo {author}
  {\bibfnamefont {G.}~\bibnamefont {Kresse}}, \bibinfo {author} {\bibfnamefont
  {T.~O.}\ \bibnamefont {Wehling}},\ and\ \bibinfo {author} {\bibfnamefont
  {M.}~\bibnamefont {Aichhorn}},\ }\href@noop {} {\bibfield  {journal}
  {\bibinfo  {journal} {Journal of Physics: Condensed Matter}\ }\textbf
  {\bibinfo {volume} {30}},\ \bibinfo {pages} {475901} (\bibinfo {year}
  {2018})}\BibitemShut {NoStop}%
\bibitem [{\citenamefont {Aichhorn}\ \emph {et~al.}(2016)\citenamefont
  {Aichhorn}, \citenamefont {Pourovskii}, \citenamefont {Seth}, \citenamefont
  {Vildosola}, \citenamefont {Zingl}, \citenamefont {Peil}, \citenamefont
  {Deng}, \citenamefont {Mravlje}, \citenamefont {Kraberger}, \citenamefont
  {Martins}, \citenamefont {Ferrero},\ and\ \citenamefont
  {Parcollet}}]{aichhorn_dfttools_2016}%
  \BibitemOpen
  \bibfield  {author} {\bibinfo {author} {\bibfnamefont {M.}~\bibnamefont
  {Aichhorn}}, \bibinfo {author} {\bibfnamefont {L.}~\bibnamefont
  {Pourovskii}}, \bibinfo {author} {\bibfnamefont {P.}~\bibnamefont {Seth}},
  \bibinfo {author} {\bibfnamefont {V.}~\bibnamefont {Vildosola}}, \bibinfo
  {author} {\bibfnamefont {M.}~\bibnamefont {Zingl}}, \bibinfo {author}
  {\bibfnamefont {O.}~\bibnamefont {Peil}}, \bibinfo {author} {\bibfnamefont
  {X.}~\bibnamefont {Deng}}, \bibinfo {author} {\bibfnamefont {J.}~\bibnamefont
  {Mravlje}}, \bibinfo {author} {\bibfnamefont {G.}~\bibnamefont {Kraberger}},
  \bibinfo {author} {\bibfnamefont {C.}~\bibnamefont {Martins}}, \bibinfo
  {author} {\bibfnamefont {M.}~\bibnamefont {Ferrero}},\ and\ \bibinfo {author}
  {\bibfnamefont {O.}~\bibnamefont {Parcollet}},\ }\href
  {https://doi.org/doi:10.1016/j.cpc.2016.03.014} {\bibfield  {journal}
  {\bibinfo  {journal} {Computer Physics Communications}\ }\textbf {\bibinfo
  {volume} {204}},\ \bibinfo {pages} {200} (\bibinfo {year}
  {2016})}\BibitemShut {NoStop}%
\bibitem [{\citenamefont {Parcollet}\ \emph {et~al.}(2015)\citenamefont
  {Parcollet}, \citenamefont {Ferrero}, \citenamefont {Ayral}, \citenamefont
  {Hafermann}, \citenamefont {Krivenko}, \citenamefont {Messio},\ and\
  \citenamefont {Seth}}]{parcollet_triqs_2015}%
  \BibitemOpen
  \bibfield  {author} {\bibinfo {author} {\bibfnamefont {O.}~\bibnamefont
  {Parcollet}}, \bibinfo {author} {\bibfnamefont {M.}~\bibnamefont {Ferrero}},
  \bibinfo {author} {\bibfnamefont {T.}~\bibnamefont {Ayral}}, \bibinfo
  {author} {\bibfnamefont {H.}~\bibnamefont {Hafermann}}, \bibinfo {author}
  {\bibfnamefont {I.}~\bibnamefont {Krivenko}}, \bibinfo {author}
  {\bibfnamefont {L.}~\bibnamefont {Messio}},\ and\ \bibinfo {author}
  {\bibfnamefont {P.}~\bibnamefont {Seth}},\ }\href
  {https://doi.org/https://doi.org/10.1016/j.cpc.2015.04.023} {\bibfield
  {journal} {\bibinfo  {journal} {Computer Physics Communications}\ }\textbf
  {\bibinfo {volume} {196}},\ \bibinfo {pages} {398 } (\bibinfo {year}
  {2015})}\BibitemShut {NoStop}%
\bibitem [{\citenamefont {Hampel}\ \emph {et~al.}(2019)\citenamefont {Hampel},
  \citenamefont {Beck},\ and\ \citenamefont {Ederer}}]{soliDMFT}%
  \BibitemOpen
  \bibfield  {author} {\bibinfo {author} {\bibfnamefont {A.}~\bibnamefont
  {Hampel}}, \bibinfo {author} {\bibfnamefont {S.}~\bibnamefont {Beck}},\ and\
  \bibinfo {author} {\bibfnamefont {C.}~\bibnamefont {Ederer}},\ }\href@noop {}
  {\bibinfo {title} {{soliDMFT}}},\ \bibinfo {howpublished}
  {\url{https://github.com/materialstheory/soliDMFT}} (\bibinfo {year}
  {2019})\BibitemShut {NoStop}%
\bibitem [{\citenamefont {Gull}\ \emph {et~al.}(2011)\citenamefont {Gull},
  \citenamefont {Millis}, \citenamefont {Lichtenstein}, \citenamefont
  {Rubtsov}, \citenamefont {Troyer},\ and\ \citenamefont {Werner}}]{Gull:2011}%
  \BibitemOpen
  \bibfield  {author} {\bibinfo {author} {\bibfnamefont {E.}~\bibnamefont
  {Gull}}, \bibinfo {author} {\bibfnamefont {A.~J.}\ \bibnamefont {Millis}},
  \bibinfo {author} {\bibfnamefont {A.~I.}\ \bibnamefont {Lichtenstein}},
  \bibinfo {author} {\bibfnamefont {A.~N.}\ \bibnamefont {Rubtsov}}, \bibinfo
  {author} {\bibfnamefont {M.}~\bibnamefont {Troyer}},\ and\ \bibinfo {author}
  {\bibfnamefont {P.}~\bibnamefont {Werner}},\ }\href
  {https://doi.org/10.1103/RevModPhys.83.349} {\bibfield  {journal} {\bibinfo
  {journal} {Reviews of Modern Physics}\ }\textbf {\bibinfo {volume} {83}},\
  \bibinfo {pages} {349} (\bibinfo {year} {2011})}\BibitemShut {NoStop}%
\bibitem [{\citenamefont {Seth}\ \emph {et~al.}(2016)\citenamefont {Seth},
  \citenamefont {Krivenko}, \citenamefont {Ferrero},\ and\ \citenamefont
  {Parcollet}}]{Seth2016274}%
  \BibitemOpen
  \bibfield  {author} {\bibinfo {author} {\bibfnamefont {P.}~\bibnamefont
  {Seth}}, \bibinfo {author} {\bibfnamefont {I.}~\bibnamefont {Krivenko}},
  \bibinfo {author} {\bibfnamefont {M.}~\bibnamefont {Ferrero}},\ and\ \bibinfo
  {author} {\bibfnamefont {O.}~\bibnamefont {Parcollet}},\ }\href
  {https://doi.org/http://dx.doi.org/10.1016/j.cpc.2015.10.023} {\bibfield
  {journal} {\bibinfo  {journal} {Computer Physics Communications}\ }\textbf
  {\bibinfo {volume} {200}},\ \bibinfo {pages} {274 } (\bibinfo {year}
  {2016})}\BibitemShut {NoStop}%
\bibitem [{\citenamefont {Vaugier}\ \emph {et~al.}(2012)\citenamefont
  {Vaugier}, \citenamefont {Jiang},\ and\ \citenamefont
  {Biermann}}]{vaugier2012}%
  \BibitemOpen
  \bibfield  {author} {\bibinfo {author} {\bibfnamefont {L.}~\bibnamefont
  {Vaugier}}, \bibinfo {author} {\bibfnamefont {H.}~\bibnamefont {Jiang}},\
  and\ \bibinfo {author} {\bibfnamefont {S.}~\bibnamefont {Biermann}},\ }\href
  {https://doi.org/10.1103/PhysRevB.86.165105} {\bibfield  {journal} {\bibinfo
  {journal} {Physical Review B}\ }\textbf {\bibinfo {volume} {86}},\ \bibinfo
  {pages} {165105} (\bibinfo {year} {2012})}\BibitemShut {NoStop}%
\bibitem [{\citenamefont {Solovyev}\ \emph {et~al.}(1994)\citenamefont
  {Solovyev}, \citenamefont {Dederichs},\ and\ \citenamefont
  {Anisimov}}]{Solovyev:1994}%
  \BibitemOpen
  \bibfield  {author} {\bibinfo {author} {\bibfnamefont {I.~V.}\ \bibnamefont
  {Solovyev}}, \bibinfo {author} {\bibfnamefont {P.~H.}\ \bibnamefont
  {Dederichs}},\ and\ \bibinfo {author} {\bibfnamefont {V.~I.}\ \bibnamefont
  {Anisimov}},\ }\href {https://doi.org/10.1103/PhysRevB.50.16861} {\bibfield
  {journal} {\bibinfo  {journal} {Physical Review B}\ }\textbf {\bibinfo
  {volume} {50}},\ \bibinfo {pages} {16861} (\bibinfo {year}
  {1994})}\BibitemShut {NoStop}%
\bibitem [{\citenamefont {Anisimov}\ \emph {et~al.}(1997)\citenamefont
  {Anisimov}, \citenamefont {Aryasetiawan},\ and\ \citenamefont
  {Lichtenstein}}]{anisimov1997}%
  \BibitemOpen
  \bibfield  {author} {\bibinfo {author} {\bibfnamefont {V.~I.}\ \bibnamefont
  {Anisimov}}, \bibinfo {author} {\bibfnamefont {F.}~\bibnamefont
  {Aryasetiawan}},\ and\ \bibinfo {author} {\bibfnamefont {A.~I.}\ \bibnamefont
  {Lichtenstein}},\ }\href {http://stacks.iop.org/0953-8984/9/i=4/a=002}
  {\bibfield  {journal} {\bibinfo  {journal} {Journal of Physics: Condensed
  Matter}\ }\textbf {\bibinfo {volume} {9}},\ \bibinfo {pages} {767} (\bibinfo
  {year} {1997})}\BibitemShut {NoStop}%
\bibitem [{\citenamefont {Hampel}\ \emph {et~al.}(2020)\citenamefont {Hampel},
  \citenamefont {Beck},\ and\ \citenamefont {Ederer}}]{Hampel2020}%
  \BibitemOpen
  \bibfield  {author} {\bibinfo {author} {\bibfnamefont {A.}~\bibnamefont
  {Hampel}}, \bibinfo {author} {\bibfnamefont {S.}~\bibnamefont {Beck}},\ and\
  \bibinfo {author} {\bibfnamefont {C.}~\bibnamefont {Ederer}},\ }\href
  {https://doi.org/10.1103/PhysRevResearch.2.033088} {\bibfield  {journal}
  {\bibinfo  {journal} {Physical Review Research}\ }\textbf {\bibinfo {volume}
  {2}},\ \bibinfo {pages} {033088} (\bibinfo {year} {2020})}\BibitemShut
  {NoStop}%
\bibitem [{\citenamefont {Haule}\ and\ \citenamefont
  {Birol}(2015)}]{PhysRevLett.115.256402}%
  \BibitemOpen
  \bibfield  {author} {\bibinfo {author} {\bibfnamefont {K.}~\bibnamefont
  {Haule}}\ and\ \bibinfo {author} {\bibfnamefont {T.}~\bibnamefont {Birol}},\
  }\href {https://doi.org/10.1103/PhysRevLett.115.256402} {\bibfield  {journal}
  {\bibinfo  {journal} {Physical Review Letters}\ }\textbf {\bibinfo {volume}
  {115}},\ \bibinfo {pages} {256402} (\bibinfo {year} {2015})}\BibitemShut
  {NoStop}%
\bibitem [{\citenamefont {Kaltak}(2015)}]{Merzuk2015}%
  \BibitemOpen
  \bibfield  {author} {\bibinfo {author} {\bibfnamefont {M.}~\bibnamefont
  {Kaltak}},\ }\emph {\bibinfo {title} {Merging GW with DMFT}},\ \href
  {http://othes.univie.ac.at/38099/} {Ph.D. thesis},\ \bibinfo  {school}
  {University of Vienna} (\bibinfo {year} {2015})\BibitemShut {NoStop}%
\bibitem [{\citenamefont {Miyake}\ \emph {et~al.}(2009)\citenamefont {Miyake},
  \citenamefont {Aryasetiawan},\ and\ \citenamefont
  {Imada}}]{PhysRevB.80.155134}%
  \BibitemOpen
  \bibfield  {author} {\bibinfo {author} {\bibfnamefont {T.}~\bibnamefont
  {Miyake}}, \bibinfo {author} {\bibfnamefont {F.}~\bibnamefont
  {Aryasetiawan}},\ and\ \bibinfo {author} {\bibfnamefont {M.}~\bibnamefont
  {Imada}},\ }\href {https://doi.org/10.1103/PhysRevB.80.155134} {\bibfield
  {journal} {\bibinfo  {journal} {Physical Review B}\ }\textbf {\bibinfo
  {volume} {80}},\ \bibinfo {pages} {155134} (\bibinfo {year}
  {2009})}\BibitemShut {NoStop}%
\bibitem [{\citenamefont {Mostofi}\ \emph {et~al.}(2014)\citenamefont
  {Mostofi}, \citenamefont {Yates}, \citenamefont {Pizzi}, \citenamefont {Lee},
  \citenamefont {Souza}, \citenamefont {Vanderbilt},\ and\ \citenamefont
  {Marzari}}]{Mostofi_et_al:2014}%
  \BibitemOpen
  \bibfield  {author} {\bibinfo {author} {\bibfnamefont {A.~A.}\ \bibnamefont
  {Mostofi}}, \bibinfo {author} {\bibfnamefont {J.~R.}\ \bibnamefont {Yates}},
  \bibinfo {author} {\bibfnamefont {G.}~\bibnamefont {Pizzi}}, \bibinfo
  {author} {\bibfnamefont {Y.-S.}\ \bibnamefont {Lee}}, \bibinfo {author}
  {\bibfnamefont {I.}~\bibnamefont {Souza}}, \bibinfo {author} {\bibfnamefont
  {D.}~\bibnamefont {Vanderbilt}},\ and\ \bibinfo {author} {\bibfnamefont
  {N.}~\bibnamefont {Marzari}},\ }\href
  {https://doi.org/https://doi.org/10.1016/j.cpc.2014.05.003} {\bibfield
  {journal} {\bibinfo  {journal} {Computer Physics Communications}\ }\textbf
  {\bibinfo {volume} {185}},\ \bibinfo {pages} {2309 } (\bibinfo {year}
  {2014})}\BibitemShut {NoStop}%
\bibitem [{\citenamefont {Park}\ \emph {et~al.}(2015)\citenamefont {Park},
  \citenamefont {Millis},\ and\ \citenamefont {Marianetti}}]{Park2015_wan}%
  \BibitemOpen
  \bibfield  {author} {\bibinfo {author} {\bibfnamefont {H.}~\bibnamefont
  {Park}}, \bibinfo {author} {\bibfnamefont {A.~J.}\ \bibnamefont {Millis}},\
  and\ \bibinfo {author} {\bibfnamefont {C.~A.}\ \bibnamefont {Marianetti}},\
  }\href {https://doi.org/10.1103/PhysRevB.92.035146} {\bibfield  {journal}
  {\bibinfo  {journal} {Physical Review B}\ }\textbf {\bibinfo {volume} {92}},\
  \bibinfo {pages} {035146} (\bibinfo {year} {2015})}\BibitemShut {NoStop}%
\bibitem [{\citenamefont {Nilsson}\ \emph {et~al.}(2017)\citenamefont
  {Nilsson}, \citenamefont {Boehnke}, \citenamefont {Werner},\ and\
  \citenamefont {Aryasetiawan}}]{Nilsson2017}%
  \BibitemOpen
  \bibfield  {author} {\bibinfo {author} {\bibfnamefont {F.}~\bibnamefont
  {Nilsson}}, \bibinfo {author} {\bibfnamefont {L.}~\bibnamefont {Boehnke}},
  \bibinfo {author} {\bibfnamefont {P.}~\bibnamefont {Werner}},\ and\ \bibinfo
  {author} {\bibfnamefont {F.}~\bibnamefont {Aryasetiawan}},\ }\href
  {https://doi.org/10.1103/PhysRevMaterials.1.043803} {\bibfield  {journal}
  {\bibinfo  {journal} {Phys. Rev. Materials}\ }\textbf {\bibinfo {volume}
  {1}},\ \bibinfo {pages} {043803} (\bibinfo {year} {2017})}\BibitemShut
  {NoStop}%
\bibitem [{\citenamefont {Petocchi}\ \emph {et~al.}(2020)\citenamefont
  {Petocchi}, \citenamefont {Nilsson}, \citenamefont {Aryasetiawan},\ and\
  \citenamefont {Werner}}]{Petocchi2020}%
  \BibitemOpen
  \bibfield  {author} {\bibinfo {author} {\bibfnamefont {F.}~\bibnamefont
  {Petocchi}}, \bibinfo {author} {\bibfnamefont {F.}~\bibnamefont {Nilsson}},
  \bibinfo {author} {\bibfnamefont {F.}~\bibnamefont {Aryasetiawan}},\ and\
  \bibinfo {author} {\bibfnamefont {P.}~\bibnamefont {Werner}},\ }\href
  {https://doi.org/10.1103/PhysRevResearch.2.013191} {\bibfield  {journal}
  {\bibinfo  {journal} {Physical Review Research}\ }\textbf {\bibinfo {volume}
  {2}},\ \bibinfo {pages} {013191} (\bibinfo {year} {2020})}\BibitemShut
  {NoStop}%
\bibitem [{\citenamefont {Kraberger}\ \emph {et~al.}(2017)\citenamefont
  {Kraberger}, \citenamefont {Triebl}, \citenamefont {Zingl},\ and\
  \citenamefont {Aichhorn}}]{Kraberger2017}%
  \BibitemOpen
  \bibfield  {author} {\bibinfo {author} {\bibfnamefont {G.~J.}\ \bibnamefont
  {Kraberger}}, \bibinfo {author} {\bibfnamefont {R.}~\bibnamefont {Triebl}},
  \bibinfo {author} {\bibfnamefont {M.}~\bibnamefont {Zingl}},\ and\ \bibinfo
  {author} {\bibfnamefont {M.}~\bibnamefont {Aichhorn}},\ }\href
  {https://doi.org/10.1103/PhysRevB.96.155128} {\bibfield  {journal} {\bibinfo
  {journal} {Physical Review B}\ }\textbf {\bibinfo {volume} {96}},\ \bibinfo
  {pages} {155128} (\bibinfo {year} {2017})}\BibitemShut {NoStop}%
\bibitem [{\citenamefont {Jarrell}\ and\ \citenamefont
  {Gubernatis}(1996)}]{Jarrel:2010}%
  \BibitemOpen
  \bibfield  {author} {\bibinfo {author} {\bibfnamefont {M.}~\bibnamefont
  {Jarrell}}\ and\ \bibinfo {author} {\bibfnamefont {J.}~\bibnamefont
  {Gubernatis}},\ }\href
  {https://doi.org/https://doi.org/10.1016/0370-1573(95)00074-7} {\bibfield
  {journal} {\bibinfo  {journal} {Physics Reports}\ }\textbf {\bibinfo {volume}
  {269}},\ \bibinfo {pages} {133 } (\bibinfo {year} {1996})}\BibitemShut
  {NoStop}%
\bibitem [{\citenamefont {Deng}\ \emph {et~al.}(2014)\citenamefont {Deng},
  \citenamefont {Sternbach}, \citenamefont {Haule}, \citenamefont {Basov},\
  and\ \citenamefont {Kotliar}}]{Xiaoyu2014}%
  \BibitemOpen
  \bibfield  {author} {\bibinfo {author} {\bibfnamefont {X.}~\bibnamefont
  {Deng}}, \bibinfo {author} {\bibfnamefont {A.}~\bibnamefont {Sternbach}},
  \bibinfo {author} {\bibfnamefont {K.}~\bibnamefont {Haule}}, \bibinfo
  {author} {\bibfnamefont {D.~N.}\ \bibnamefont {Basov}},\ and\ \bibinfo
  {author} {\bibfnamefont {G.}~\bibnamefont {Kotliar}},\ }\href
  {https://doi.org/10.1103/PhysRevLett.113.246404} {\bibfield  {journal}
  {\bibinfo  {journal} {Physical Review Letters}\ }\textbf {\bibinfo {volume}
  {113}},\ \bibinfo {pages} {246404} (\bibinfo {year} {2014})}\BibitemShut
  {NoStop}%
\bibitem [{\citenamefont {Ferber}\ \emph {et~al.}(2012)\citenamefont {Ferber},
  \citenamefont {Foyevtsova}, \citenamefont {Valent\'{\i}},\ and\ \citenamefont
  {Jeschke}}]{Ferber2012}%
  \BibitemOpen
  \bibfield  {author} {\bibinfo {author} {\bibfnamefont {J.}~\bibnamefont
  {Ferber}}, \bibinfo {author} {\bibfnamefont {K.}~\bibnamefont {Foyevtsova}},
  \bibinfo {author} {\bibfnamefont {R.}~\bibnamefont {Valent\'{\i}}},\ and\
  \bibinfo {author} {\bibfnamefont {H.~O.}\ \bibnamefont {Jeschke}},\ }\href
  {https://doi.org/10.1103/PhysRevB.85.094505} {\bibfield  {journal} {\bibinfo
  {journal} {Physical Review B}\ }\textbf {\bibinfo {volume} {85}},\ \bibinfo
  {pages} {094505} (\bibinfo {year} {2012})}\BibitemShut {NoStop}%
\bibitem [{\citenamefont {Amadon}\ \emph {et~al.}(2006)\citenamefont {Amadon},
  \citenamefont {Biermann}, \citenamefont {Georges},\ and\ \citenamefont
  {Aryasetiawan}}]{Amadon2006}%
  \BibitemOpen
  \bibfield  {author} {\bibinfo {author} {\bibfnamefont {B.}~\bibnamefont
  {Amadon}}, \bibinfo {author} {\bibfnamefont {S.}~\bibnamefont {Biermann}},
  \bibinfo {author} {\bibfnamefont {A.}~\bibnamefont {Georges}},\ and\ \bibinfo
  {author} {\bibfnamefont {F.}~\bibnamefont {Aryasetiawan}},\ }\href
  {https://doi.org/10.1103/PhysRevLett.96.066402} {\bibfield  {journal}
  {\bibinfo  {journal} {Physical Review Letters}\ }\textbf {\bibinfo {volume}
  {96}},\ \bibinfo {pages} {066402} (\bibinfo {year} {2006})}\BibitemShut
  {NoStop}%
\bibitem [{\citenamefont {Dang}\ \emph {et~al.}(2015)\citenamefont {Dang},
  \citenamefont {Mravlje}, \citenamefont {Georges},\ and\ \citenamefont
  {Millis}}]{Dang2015}%
  \BibitemOpen
  \bibfield  {author} {\bibinfo {author} {\bibfnamefont {H.~T.}\ \bibnamefont
  {Dang}}, \bibinfo {author} {\bibfnamefont {J.}~\bibnamefont {Mravlje}},
  \bibinfo {author} {\bibfnamefont {A.}~\bibnamefont {Georges}},\ and\ \bibinfo
  {author} {\bibfnamefont {A.~J.}\ \bibnamefont {Millis}},\ }\href
  {https://doi.org/10.1103/PhysRevB.91.195149} {\bibfield  {journal} {\bibinfo
  {journal} {Physical Review B}\ }\textbf {\bibinfo {volume} {91}},\ \bibinfo
  {pages} {195149} (\bibinfo {year} {2015})}\BibitemShut {NoStop}%
\bibitem [{\citenamefont {Kim}\ and\ \citenamefont {Min}(2015)}]{Kim2015}%
  \BibitemOpen
  \bibfield  {author} {\bibinfo {author} {\bibfnamefont {M.}~\bibnamefont
  {Kim}}\ and\ \bibinfo {author} {\bibfnamefont {B.~I.}\ \bibnamefont {Min}},\
  }\href {https://doi.org/10.1103/PhysRevB.91.205116} {\bibfield  {journal}
  {\bibinfo  {journal} {Physical Review B}\ }\textbf {\bibinfo {volume} {91}},\
  \bibinfo {pages} {205116} (\bibinfo {year} {2015})}\BibitemShut {NoStop}%
\bibitem [{\citenamefont {Schäfer}\ \emph {et~al.}(2020)\citenamefont
  {Schäfer}, \citenamefont {Wentzell}, \citenamefont {Šimkovic},
  \citenamefont {He}, \citenamefont {Hille}, \citenamefont {Klett},
  \citenamefont {Eckhardt}, \citenamefont {Arzhang}, \citenamefont {Harkov},
  \citenamefont {Régent}, \citenamefont {Kirsch}, \citenamefont {Wang},
  \citenamefont {Kim}, \citenamefont {Kozik}, \citenamefont {Stepanov},
  \citenamefont {Kauch}, \citenamefont {Andergassen}, \citenamefont {Hansmann},
  \citenamefont {Rohe}, \citenamefont {Vilk}, \citenamefont {LeBlanc},
  \citenamefont {Zhang}, \citenamefont {Tremblay}, \citenamefont {Ferrero},
  \citenamefont {Parcollet},\ and\ \citenamefont {Georges}}]{schaefer2020}%
  \BibitemOpen
  \bibfield  {author} {\bibinfo {author} {\bibfnamefont {T.}~\bibnamefont
  {Schäfer}}, \bibinfo {author} {\bibfnamefont {N.}~\bibnamefont {Wentzell}},
  \bibinfo {author} {\bibfnamefont {F.}~\bibnamefont {Šimkovic}}, \bibinfo
  {author} {\bibfnamefont {Y.-Y.}\ \bibnamefont {He}}, \bibinfo {author}
  {\bibfnamefont {C.}~\bibnamefont {Hille}}, \bibinfo {author} {\bibfnamefont
  {M.}~\bibnamefont {Klett}}, \bibinfo {author} {\bibfnamefont {C.~J.}\
  \bibnamefont {Eckhardt}}, \bibinfo {author} {\bibfnamefont {B.}~\bibnamefont
  {Arzhang}}, \bibinfo {author} {\bibfnamefont {V.}~\bibnamefont {Harkov}},
  \bibinfo {author} {\bibfnamefont {F.-M.~L.}\ \bibnamefont {Régent}},
  \bibinfo {author} {\bibfnamefont {A.}~\bibnamefont {Kirsch}}, \bibinfo
  {author} {\bibfnamefont {Y.}~\bibnamefont {Wang}}, \bibinfo {author}
  {\bibfnamefont {A.~J.}\ \bibnamefont {Kim}}, \bibinfo {author} {\bibfnamefont
  {E.}~\bibnamefont {Kozik}}, \bibinfo {author} {\bibfnamefont {E.~A.}\
  \bibnamefont {Stepanov}}, \bibinfo {author} {\bibfnamefont {A.}~\bibnamefont
  {Kauch}}, \bibinfo {author} {\bibfnamefont {S.}~\bibnamefont {Andergassen}},
  \bibinfo {author} {\bibfnamefont {P.}~\bibnamefont {Hansmann}}, \bibinfo
  {author} {\bibfnamefont {D.}~\bibnamefont {Rohe}}, \bibinfo {author}
  {\bibfnamefont {Y.~M.}\ \bibnamefont {Vilk}}, \bibinfo {author}
  {\bibfnamefont {J.~P.~F.}\ \bibnamefont {LeBlanc}}, \bibinfo {author}
  {\bibfnamefont {S.}~\bibnamefont {Zhang}}, \bibinfo {author} {\bibfnamefont
  {A.~M.~S.}\ \bibnamefont {Tremblay}}, \bibinfo {author} {\bibfnamefont
  {M.}~\bibnamefont {Ferrero}}, \bibinfo {author} {\bibfnamefont
  {O.}~\bibnamefont {Parcollet}},\ and\ \bibinfo {author} {\bibfnamefont
  {A.}~\bibnamefont {Georges}},\ }\href@noop {} {} (\bibinfo {year} {2020}),\
  \Eprint {https://arxiv.org/abs/2006.10769} {arXiv:2006.10769} \BibitemShut
  {NoStop}%
\bibitem [{\citenamefont {Haule}(2015)}]{Haule:2015_exactDC}%
  \BibitemOpen
  \bibfield  {author} {\bibinfo {author} {\bibfnamefont {K.}~\bibnamefont
  {Haule}},\ }\href {https://doi.org/10.1103/PhysRevLett.115.196403} {\bibfield
   {journal} {\bibinfo  {journal} {Physical Review Letters}\ }\textbf {\bibinfo
  {volume} {115}},\ \bibinfo {pages} {196403} (\bibinfo {year}
  {2015})}\BibitemShut {NoStop}%
\bibitem [{\citenamefont {Kristanovski}\ \emph {et~al.}(2018)\citenamefont
  {Kristanovski}, \citenamefont {Shick}, \citenamefont {Lechermann},\ and\
  \citenamefont {Lichtenstein}}]{Kristanovski:2018}%
  \BibitemOpen
  \bibfield  {author} {\bibinfo {author} {\bibfnamefont {O.}~\bibnamefont
  {Kristanovski}}, \bibinfo {author} {\bibfnamefont {A.~B.}\ \bibnamefont
  {Shick}}, \bibinfo {author} {\bibfnamefont {F.}~\bibnamefont {Lechermann}},\
  and\ \bibinfo {author} {\bibfnamefont {A.~I.}\ \bibnamefont {Lichtenstein}},\
  }\href {https://doi.org/10.1103/PhysRevB.97.201116} {\bibfield  {journal}
  {\bibinfo  {journal} {Phys. Rev. B}\ }\textbf {\bibinfo {volume} {97}},\
  \bibinfo {pages} {201116} (\bibinfo {year} {2018})}\BibitemShut {NoStop}%
\bibitem [{\citenamefont {Nekrasov}\ \emph {et~al.}(2012)\citenamefont
  {Nekrasov}, \citenamefont {Pavlov},\ and\ \citenamefont
  {Sadovskii}}]{Nekrasov2012}%
  \BibitemOpen
  \bibfield  {author} {\bibinfo {author} {\bibfnamefont {I.~A.}\ \bibnamefont
  {Nekrasov}}, \bibinfo {author} {\bibfnamefont {V.~S.}\ \bibnamefont
  {Pavlov}},\ and\ \bibinfo {author} {\bibfnamefont {M.~V.}\ \bibnamefont
  {Sadovskii}},\ }\href {https://doi.org/10.1134/S0021364012110070} {\bibfield
  {journal} {\bibinfo  {journal} {JETP Letters}\ }\textbf {\bibinfo {volume}
  {95}},\ \bibinfo {pages} {581} (\bibinfo {year} {2012})}\BibitemShut
  {NoStop}%
\bibitem [{\citenamefont {Georges}\ \emph {et~al.}(1996)\citenamefont
  {Georges}, \citenamefont {Kotliar}, \citenamefont {Krauth},\ and\
  \citenamefont {Rozenberg}}]{Georges:1996}%
  \BibitemOpen
  \bibfield  {author} {\bibinfo {author} {\bibfnamefont {A.}~\bibnamefont
  {Georges}}, \bibinfo {author} {\bibfnamefont {G.}~\bibnamefont {Kotliar}},
  \bibinfo {author} {\bibfnamefont {W.}~\bibnamefont {Krauth}},\ and\ \bibinfo
  {author} {\bibfnamefont {M.~J.}\ \bibnamefont {Rozenberg}},\ }\href
  {https://doi.org/10.1103/RevModPhys.68.13} {\bibfield  {journal} {\bibinfo
  {journal} {Reviews of Modern Physics}\ }\textbf {\bibinfo {volume} {68}},\
  \bibinfo {pages} {13} (\bibinfo {year} {1996})}\BibitemShut {NoStop}%
\bibitem [{\citenamefont {Berthod}\ \emph {et~al.}(2013)\citenamefont
  {Berthod}, \citenamefont {Mravlje}, \citenamefont {Deng}, \citenamefont
  {\ifmmode~\check{Z}\else \v{Z}\fi{}itko}, \citenamefont {van~der Marel},\
  and\ \citenamefont {Georges}}]{Berthod:2013}%
  \BibitemOpen
  \bibfield  {author} {\bibinfo {author} {\bibfnamefont {C.}~\bibnamefont
  {Berthod}}, \bibinfo {author} {\bibfnamefont {J.}~\bibnamefont {Mravlje}},
  \bibinfo {author} {\bibfnamefont {X.}~\bibnamefont {Deng}}, \bibinfo {author}
  {\bibfnamefont {R.}~\bibnamefont {\ifmmode~\check{Z}\else \v{Z}\fi{}itko}},
  \bibinfo {author} {\bibfnamefont {D.}~\bibnamefont {van~der Marel}},\ and\
  \bibinfo {author} {\bibfnamefont {A.}~\bibnamefont {Georges}},\ }\href
  {https://doi.org/10.1103/PhysRevB.87.115109} {\bibfield  {journal} {\bibinfo
  {journal} {Physical Review B}\ }\textbf {\bibinfo {volume} {87}},\ \bibinfo
  {pages} {115109} (\bibinfo {year} {2013})}\BibitemShut {NoStop}%
\end{thebibliography}%


\end{document}